\RequirePackage{fix-cm}
\documentclass[fleqn,usenatbib]{mnras}

\usepackage{newtxtext,newtxmath}
\usepackage[T1]{fontenc}
\usepackage{ae,aecompl}

\usepackage{graphicx}	

\title[Off-axis emission of GW-detected short GRBs]{Off-axis emission of
  short $\gamma$-ray bursts and the detectability of electromagnetic
  counterparts of gravitational wave detected binary mergers}

\author[Lazzati et al.]{
Davide Lazzati$^1$, Alex Deich$^2$, Brian J. Morsony$^3$, and Jared C. Workman$^4$\\
$^1$ Department of Physics, 301 Weniger Hall, Oregon State University,
Corvallis, OR 97331\\
$^2$ Reed College, 3203 Southeast Woodstock Boulevard Portland, Oregon 97202-8199\\
$^3$ Department of Astronomy, University of Maryland, 1113 Physical
Sciences Complex, 
College Park, MD 20742-2421, USA\\
$^4$ Department of Physical and Environmental Sciences, Colorado Mesa
University, 
Grand Junction, CO 81501, USA}

\begin{document}
\label{firstpage}
\pagerange{\pageref{firstpage}--\pageref{lastpage}}
\maketitle

\begin{abstract}
  We present calculations of the wide angle emission of short-duration
  gamma-ray bursts from compact binary merger progenitors. Such events
  are expected to be localized by their gravitational wave emission,
  fairly irrespective of the orientation of the angular momentum
  vector of the system, along which the gamma-ray burst outflow is
  expected to propagate. We show that both the prompt and afterglow
  emission are dim and challenging to detect for observers lying
  outside of the cone within which the relativistic outflow is
  propagating. If the jet initially propagates through a baryon
  contaminated region surrounding the merger site, however, a hot
  cocoon forms around it. The cocoon subsequently expands
  quasi-isotropically producing its own prompt emission and external
  shock powered afterglow. We show that the cocoon prompt emission is
  detectable by Swift BAT and Fermi GBM, We also show that the cocoon
  afterglow peaks a few hours to a few days after the burst and is
  detectable for up to a few weeks at all wavelengths. The timing and
  brightness of the transient are however uncertain due to their
  dependence on unknown quantities such as the density of the ambient
  medium surrounding the merger site, the cocoon energy, and the
  cocoon Lorentz factor. For a significant fraction of the
  gravitationally-detected neutron-star-binary mergers, the cocoon
  afterglow could possibly be the only identifiable electromagnetic
  counterpart, at least at radio and X-ray frequencies.
\end{abstract}




\section{Introduction}

Short duration gamma-ray bursts (SGRBs) are expected to be associated
with the merger of a compact binary system in which at least one of
the two components is a neutron star (NS), the other possibly being a
black hole (BH;
\citealt{Eichleretal1989,Nakar2007,Berger2014}). Binary
NS\footnote{Here and in the following we call binary NS a system in
  which at least one of the components is a NS, the other being either
  a NS or a BH.}  mergers are also candidate sources of gravitational
waves, similar to the binary BH mergers recently detected by LIGO
\citep{Phinney1991,ShibataUry2002,FaberRasio2002,Sekiguchietal2011,Abbottetal2016a,Abbottetal2016b}.
While a SGRB is highly beamed and its detection against the gamma-ray
background is possible only if its relativistic jet is pointing
towards the Earth, gravitational waves are only moderately aspherical
and NS binary mergers are expected to be detectable irrespective of
their orientation. For a typical SGRB jet opening angle of $16^\circ$
\citep{Fongetal2015}, the probability of detecting an on-axis burst is
only $\sim10$ per cent\footnote{Here and in the following, these
  estimates include the angle dependency of the GW strain, which is
  larger for a face-on binary system, likely producing an on-axis
  burst \citep{Abadieetal2012}}. In other words, only one out of 10 NS
binary mergers detected in gravitational waves (GWs) will be
accompanied by a full-fledged SGRB. Other estimates put the average
opening angle at $\sim6^\circ$ \citep{Ghirlandaetal2016}, with a rate
of only one on-axis burst every $\sim100$ detected by LIGO. The most
likely orientation will be the one most difficult to observe in
electromagnetic waves, i.e., the edge-on configuration with the
relativistic jet expanding perpendicularly to the line of sight. Our
hope of confirming the supposed association of SGRBs with NS binary
mergers
\citep{ShibataTaniguchi2006,Faberetal2006,Giacomazzoetal2013,FongBerger2013,Ruizetal2016}
is therefore tied to our capability of modeling and detecting the
off-axis emission of such events \citep{MetzgerBerger2012}. While said
emission is faint, compared to the burst of gamma rays from an on-axis
jet, LIGO only detects NS binary mergers within a relatively small
distance form Earth ($\sim200$~Mpc), and even the faint off-axis SGRB
components might be detectable.

Off-axis emission from relativistic jets in the external shock phase
has been studied for long and short duration GRBs, especially for the
late afterglow phase
\citep{Rhoads1999,Granotetal2002,Rossietal2004,Rossietal2008,Eertenetal2010,Salafiaetal2015,Salafiaetal2016}. While
such off-axis events have been searched in multi-wavelength surveys,
no credible candidate has emerged, so far
\citep{Greineretal2000,Nakaretal2002,RauGreiner2005,Rykoffetal2005,Rauetal2006,Guidorzietal2009,Ghirlandaetal2015}. The
case of GW-detected NS binary mergers would, however, be different. If
a localization could be made for the direction to the GW event,
searching for a transient in a small area of the sky would be much
easier than blindly surveying a large region of the sky. Detection of
off-axis emission from SGRBs associated with gravitationally detected
binary NS mergers could then become a powerful tool for studying the
structure of the jet and of any non-relativistic or mildly
relativistic ejecta associated with the merger and the burst.

In light of these considerations, we present in this paper a
calculation of possible components of off-axis emission from a SGRB,
including the de-beamed prompt emission, the emission associated with
the jet cocoon (\cite{RamirezRuizetal2002,NakarPiran2017} who,
however, considered mainly the cocoons of long duration GRBs), and the
afterglow emission. This paper is organized as follows: in Section~2
we present the theoretical framework at the base of our calculations,
in Section~3 we present our multi-wavelength results, and in Section~4
we discuss the detectability of the signal for expected distances of
LIGO-detected NS binary mergers and the implications of a
(non)detection.

\section{Emission components}

\begin{table*}
\centering
\caption{List of symbols used and their meaning. Fiducial values are
  given for primary quantities. Variables for which a fiducial value
  is not reported can be derived from the primary ones. The
  calculated fluxes at one day for the on-axis SGRB with our fiducial
  properties, for a source at $z=0.5$, are $f_{1\rm{keV}}=0.05$~$\mu$Jy;
  $R_{AB}=22$, and $f_{8GHz}=100$~$\mu$Jy. 
  All these fluxes are within the distributions of the observed
  values as reported in Figure~1
  of~\citet{Fongetal2015}, making our fiducial values adecuate. The
  optical magnitude is in the brighter end of the observed
  distribution, and our optical band estimates may therefore be
  slightly optimistic (see also Figure 2 of \citealt{Lietal2016}).}
\label{tab:symbols}
\begin{tabular}{lll} 
  \hline
  Symbol & Fiducial value & Meaning \\ \hline
  $E_2$ & $2\times10^{50}$ erg & Total energy released by the engine in two jets \\
  $E_1$ & & Total energy released by the engine in one jet \\
  $L_2$ & & Engine luminosity in two jets \\
  $E_{\rm{iso}}$ & & Isotropic equivalent energy released by the engine \\
  $t_{\rm{eng}}$ & 1 s& Time during which the engine is active \\
  $\theta_j$ & 16$^\circ$ & Half-opening angle of each jet \\
  $\Omega_2$ && Solid angle occupied by the two jets\\
  $\Gamma_0$ & 1 & Lorentz factor of the jet at the injection
                   radius $r_0$\\
  $\Gamma_\infty$ & 100 & Maximum Lorentz factor the jet can attain\\
  $m_0$ && Rest mass of the outflow\\
  $r_0$ & $10^{7}$ cm & Radius at which the jet is launched \\
  $\theta_{\rm{obs}}$ & & Angle between the jet axis and the line of
                          sight\\
  $\theta_{v,\rm{obs}}$ & & Angle between the line of sight and the
                            velocity vector of an outflow element\\
  $\theta_{\rm{v}}$ & & Angle between the 
                        velocity vector of an outflow element and
                        the jet axis\\
  $\beta_{\rm{h}}$ & & Velocity of the head of the jet in units of the
                       speed of light\\
  $\beta_{\rm{j}}$ & & Velocity of the jet material in units of the
                       speed of light\\
  $R_{\rm{a}}$ & $10^{8}$ cm & Radius of the sphere polluted by NR ejecta\\
  $\rho_{\rm{a}}$ & $10^{7}$ g/cm$^3$ & Density of the sphere polluted by NR ejecta\\
  $R_{\rm{rad}}$ & $10^{13}$ cm & Radius at which the fireball
                                  begins releasing the prompt
                                  emission photons\\
  $L_{\rm{prompt,pk}}$ &  & Peak bolometric luminosity of the prompt emission\\
  $\Sigma$ &  & Emitting surface of the SGRB outflow\\
  $\delta$ &  & Doppler factor\\
  $\alpha_{\rm{ph}}$ & 0  & Low-frequency photon index of the prompt Band spectrum\\
  $\beta_{\rm{ph}}$ & -2.5  & High-frequency photon index of the prompt Band spectrum\\
  $h\nu_{\rm{pk}}^\prime$ & 2.5 keV  & Comoving peak photon energy of
                                       the prompt Band spectrum\\
  $\epsilon_e$ & 0.1 & Fraction of the external shock energy in
                       non-thermal electrons \\
  $\epsilon_B$ & 0.01 & Fraction of the external shock energy in
                        magnetic field \\
  $n_{\rm{ISM}}$ & 0.1 cm$^{-3}$ & density of the ambient medium for
                                   the afterglow calculations \\
  $p$ & 2.5& Index of the non-thermal electron energy distribution ($p(\gamma)\propto\gamma^{-p}$)\\ 
  $\delta{t_{\rm{rad}}^\prime}$ & 200 s  & Comoving duration of the
                                           prompt emission\\
  $t_{\rm{bo}}$ &  & Breakout time of the jet off the ambient material\\
  $E_{\rm{c}}$ &  $10^{49}$ erg& Cocoon energy\\
 $L_{\rm{c}}$ & & Prompt emission luminosity of the cocoon \\
  $\Sigma_{\rm{j}}$ &  & Cross-sectional area of the two jets at break-out\\
  $\Gamma_{\infty,{\rm c}}$ &  10 & Asymptotic Lorentz factor of the
                                    cocoon material\\
  $R_{\rm{sat,c}}$ &  & Saturation radius of the cocoon\\
  $R_{\rm{ph,c}}$ &  & Photospheric radius of the cocoon\\
  $T_{\rm{0,c}}$ &  & Initial temperature of the cocoon at breakout time\\
  $T_{\rm{ph,c}}$ &  10~keV & Observed photospheric temperature of the cocoon\\
  $V_{\rm{c}}$ &  & Volume of the cocoon at breakout time\\
  $\delta t_{\rm{diff}}$ &  & Photon diffusion time in the outflow at
                              the photosphere\\
  $\delta t_{\rm{ang}}$ &  & Angular timescale of the fireball at the photosphere\\
  \hline
\end{tabular}
\end{table*}

We consider a SGRB engine releasing a total energy $E_2$ in two
counter-propagating jets.  $E_1=E_2/2$ is the energy released in each
jet. The engine is active for a time $t_{\rm{eng}}$ and the outflow is
initially beamed in jets with an half-opening angle $\theta_j$.  The
jets are released at rest ($\Gamma_0=1$) at a distance $r_0$ from the
center of the system with a ratio of internal energy to rest mass
allowing for acceleration to a maximum Lorentz factor $\Gamma_\infty$.  The
isotropic equivalent energy of the system is
$E_{\rm{iso}}=4\pi E_2/\Omega_2$, where
$\Omega_2=4\pi(1-\cos\theta_j)$ is the solid angle occupied by the two
jets. The rest mass carried by the fireball is
$m_0=E_2/\Gamma_0{}c^2$.  We consider to counter-propagating
``top-hat'' jets, with uniform properties within their opening angle
$\theta_j$ and with sharp edges.

Guided by binary merger simulation results
\citep{Kiuchietal2014,Kiuchietal2015,Radiceetal2016} we assume that
the jet is launched and initially propagates within a baryon
contaminated environment. We assume the polluted region to be of size
$R_a$ and to have uniform density $\rho_a$. Due to the very short
interaction time with the expanding jet, we neglect any expansion of
the polluted region, which is predicted to be moving outward at
non-relativistic (NR) speed.

\subsection{Prompt emission}

\begin{figure}
\includegraphics[width=\columnwidth]{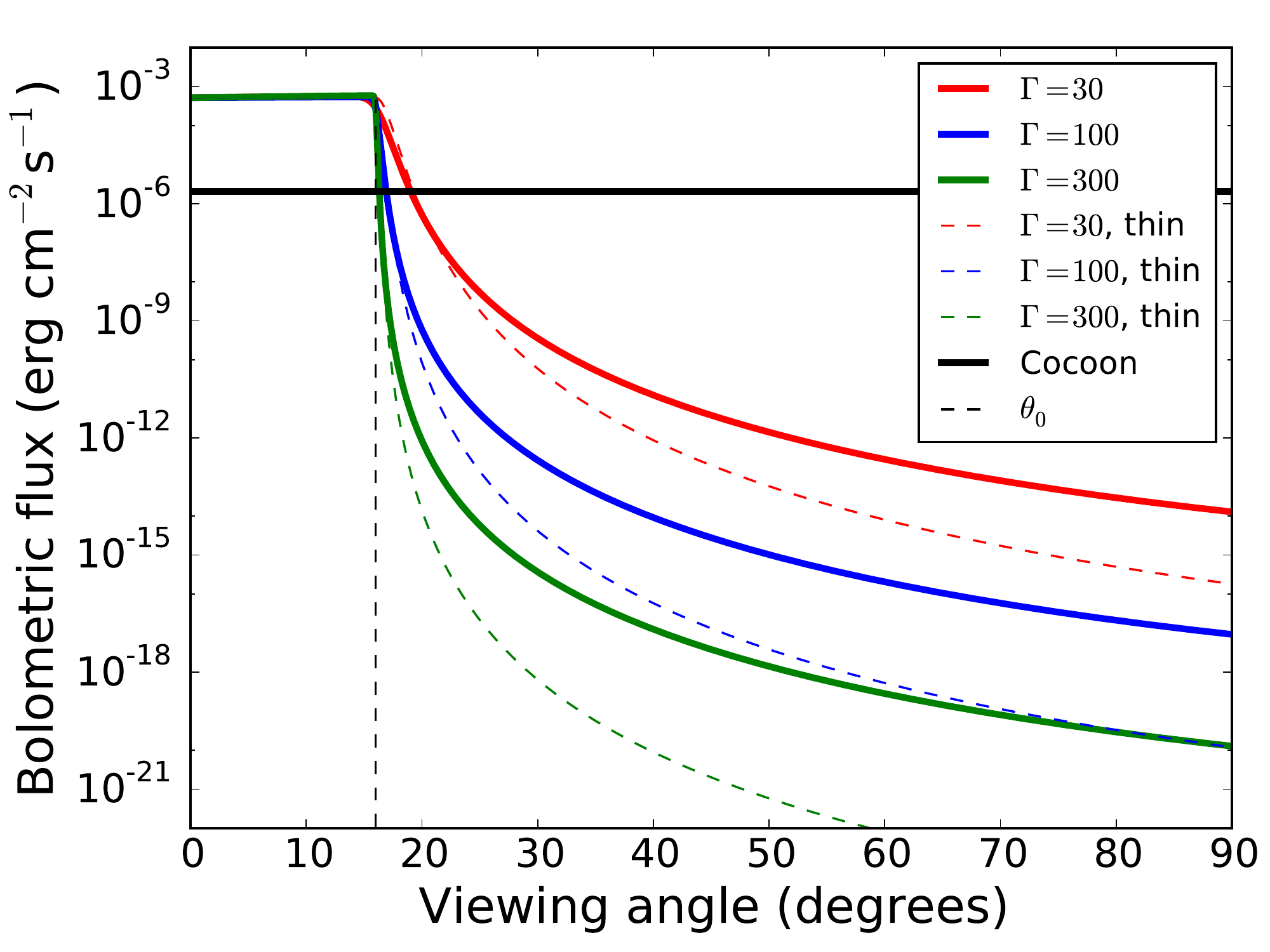}
\caption{{Bolometric peak flux of the prompt emission of a short GRB
    as a function of the observer viewing angle. The burst
    characteristics are from the fiducial values in
    Table~\ref{tab:symbols} and is located at a distance of 200~Mpc
    from Earth. Three values of the Lorentz factor are shown, as well
    as the isotropic cocoon contribution. The thin dashed line shows
    the result for a radially thin outflow that emits for a
    vanishingly small time.}
\label{fig:prompt}}
\end{figure}

The prompt emission of the SGRB jet is computed as follows (see also
\cite{Yamazakietal2002,Yamazakietal2003a,Yamazakietal2003b} for
analogous calculations for long GRBs). We assume that at a certain
distance $R_{\rm{rad}}$ from the engine the internal energy of the
outflow is converted into photons and radiated. The nature of the
emission mechanism is not discussed here and is irrelevant for our
conclusions. We assume that there are no small-scale relativistic
internal motions such as those predicted by turbulence-driven models
\citep{NarayanKumar2009} and some magnetic reconnection models
\citep{ZhangYan2011}. We also assume that the local dissipation rate
in the outflow comoving frame is uniform within the jet opening angle
and that the emitted radiation is isotropic in the local comoving
frame.

We first consider the peak bolometric luminosity $L_{\rm{prompt,pk}}$,
so that the details of the spectral shape of the prompt emission do
not need to be discussed. Its variation for observers along different
lines of sight depends on the timing properties of the jet
emission. We first consider the case in which the duration of the
emission episode in the comoving frame is long enough that every
observer sees emission from the entire outflow, at least at some
time. Under such conditions, the received bolometric flux is the
result of the integration over the emitting surface of the local
emission boosted by the Doppler factor
$\delta(\Gamma,\theta)=\left[\Gamma\left(1-\beta\cos\theta\right)\right]^{-1}$
elevated to the fourth power (one for the photon rate, one for the
blueshift in the photon energy, and two for the beaming of the solid
angle). The observed peak bolometric luminosity is therefore
calculated as:
\begin{equation}
L_{\rm{prompt,pk}}\left(\theta_{\rm{obs}}\right)=L_{\rm{prompt,pk}}(0)
\frac{\int_\Sigma\delta^4\left(\Gamma_0,\theta_{v,\rm{obs}}\right)d\sigma}
{\int_\Sigma\delta^4\left(\Gamma_0,\theta_v\right)d\sigma}
\label{eq:prompt}
\end{equation}
where $\theta_{\rm{obs}}$ is the angle between the line of sight and
the jet axis that points nearest to the observer, $\Sigma$ is the
emitting surface, $\theta_{v,\rm{obs}}$ is the angle between the line
of sight and the local velocity vector, $\theta_{v}$ is the angle
between the local velocity vector and the jet axis, and
$L_{\rm{prompt,pk}}(0)$ is the peak luminosity seen by an on-axis
observer.  The results of the above integration for various values of
$\Gamma_0$ are shown in Figure~\ref{fig:prompt}. The main lesson from
the figure is that, for expected values of the fireball Lorentz factor
at the time of the prompt emission, the observed flux drops-off quite
dramatically as soon as the line of sight moves out of the jet.

We note that the assumptions above are quite optimistic. Let us
consider now the opposite assumption, i.e., a jet in which the emission
lasts for a negligibly small amount of time in the comoving frame. In
this case each observer sees only a small region of the jet active at
any time, and the peak flux is due solely to the region of the
fireball closer to the line of sight. Eq.~\ref{eq:prompt} is replaced
by the simpler:
\begin{eqnarray}
L_{\rm{prompt,pk}}\left(\theta_{\rm{obs}}\right)&=&L_{\rm{prompt,pk}}(0)
\frac{\delta^4\left(\Gamma_0,\theta_{v,\rm{obs},\min}\right)}{\delta^4\left(\Gamma_0,\theta_v\right)}\nonumber
  \\
&\sim&\left\{
\begin{array}{ll}
L_{\rm{prompt,pk}}(0) & \theta_{\rm{obs}}\le\theta_j \\
\frac{L_{\rm{prompt,pk}}(0)}{16\Gamma_0^8\left[1-\beta\cos(\theta_{\rm{obs}}-\theta_j)\right]^4}
                   & \theta_{\rm{obs}}>\theta_j
\end{array}
\right.
\label{eq:promptthin}
\end{eqnarray}
which decays even faster than the one from Equation~\ref{eq:prompt}
for large off-axis angles (see dashed lines in Figure~\ref{fig:prompt}).

\begin{figure}
\includegraphics[width=\columnwidth]{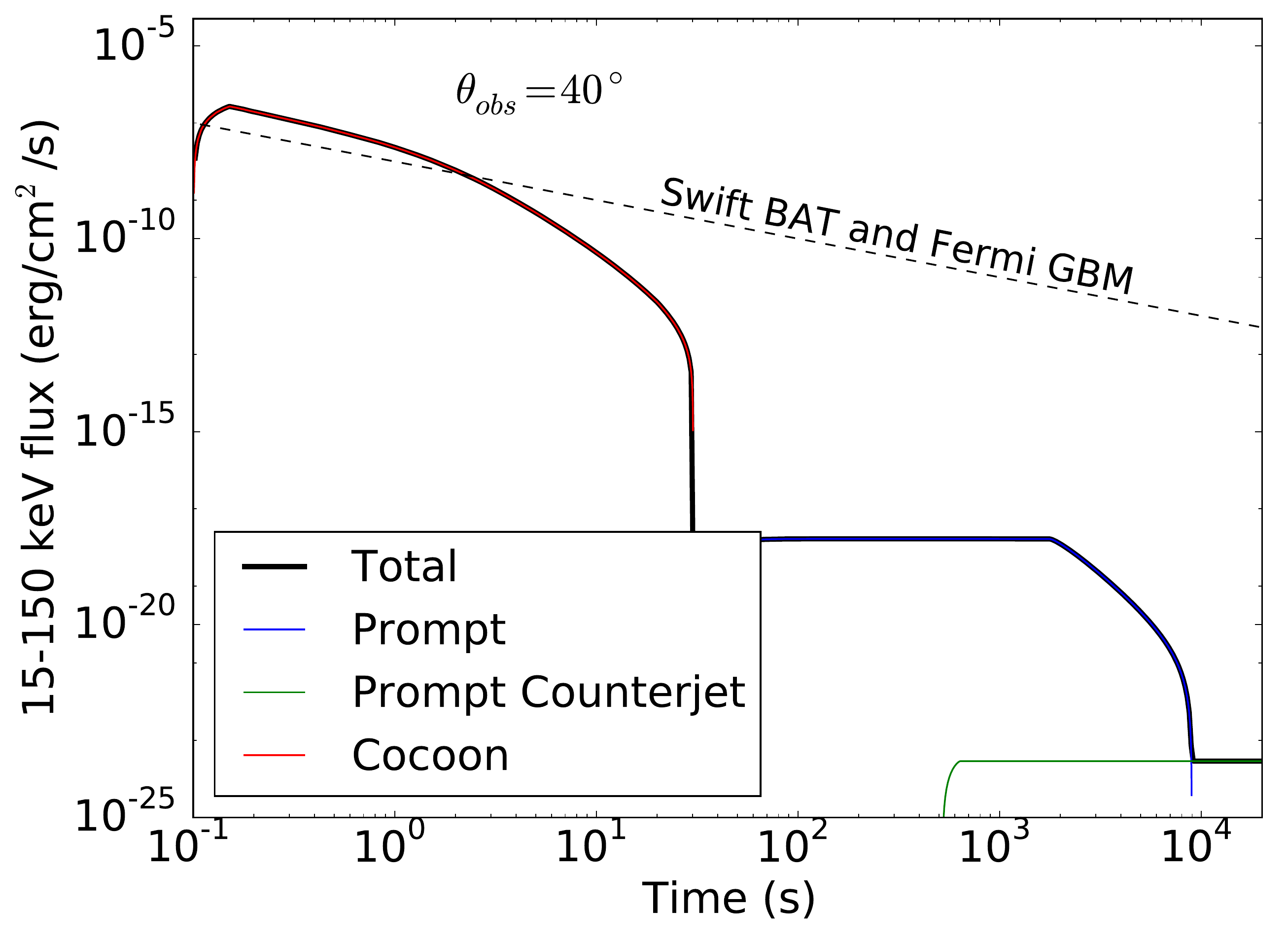}
\caption{{Prompt emission from a short GRB seen at $40^\circ$ off-axis
  in the Swift BAT or Fermi-GBM bands. The thresholds of the two
  instruments, which are similar for a low-temperature thermal source,
  are shown for comparison.}
\label{fig:coc_prompt}}
\end{figure}
 
For the calculation of the light curves in the radio, optical, X-ray,
and Swift BAT band, instead, we proceeded as follows. We assumed that
the prompt emission has a Band spectral shape with
$\alpha_{\rm{ph}}=0$, $\beta_{\rm{ph}}=-2.5$, and comoving peak
frequency $h\nu_{\rm{pk}}^\prime=2.5$~keV. We also assumed that the
emission of radiation turns on at the same radius $R_{\rm{rad}}$ for
the entire fireball and lasts, in the comoving frame
$\delta{t_{\rm{rad}}^\prime}=200$~s.  As seen from an on-axis observer
($\theta_{\rm{obs}}=0$) and assuming $\Gamma_\infty=100$, this burst
has an observed peak photon energy $h\nu_{\rm{pk}}\sim500$~keV and a
duration $\delta{t_{\rm{rad}}}\sim1$~s, fairly typical for observed
SGRBs. The spectrum was normalized such that the on-axis observer
would detect a bolometric isotropic equivalent energy
$2.5\times10^{51}$ erg, corresponding to a prompt emission efficiency
of 50 per cent.

\begin{figure*}
\parbox{\columnwidth}{
\includegraphics[width=\columnwidth]{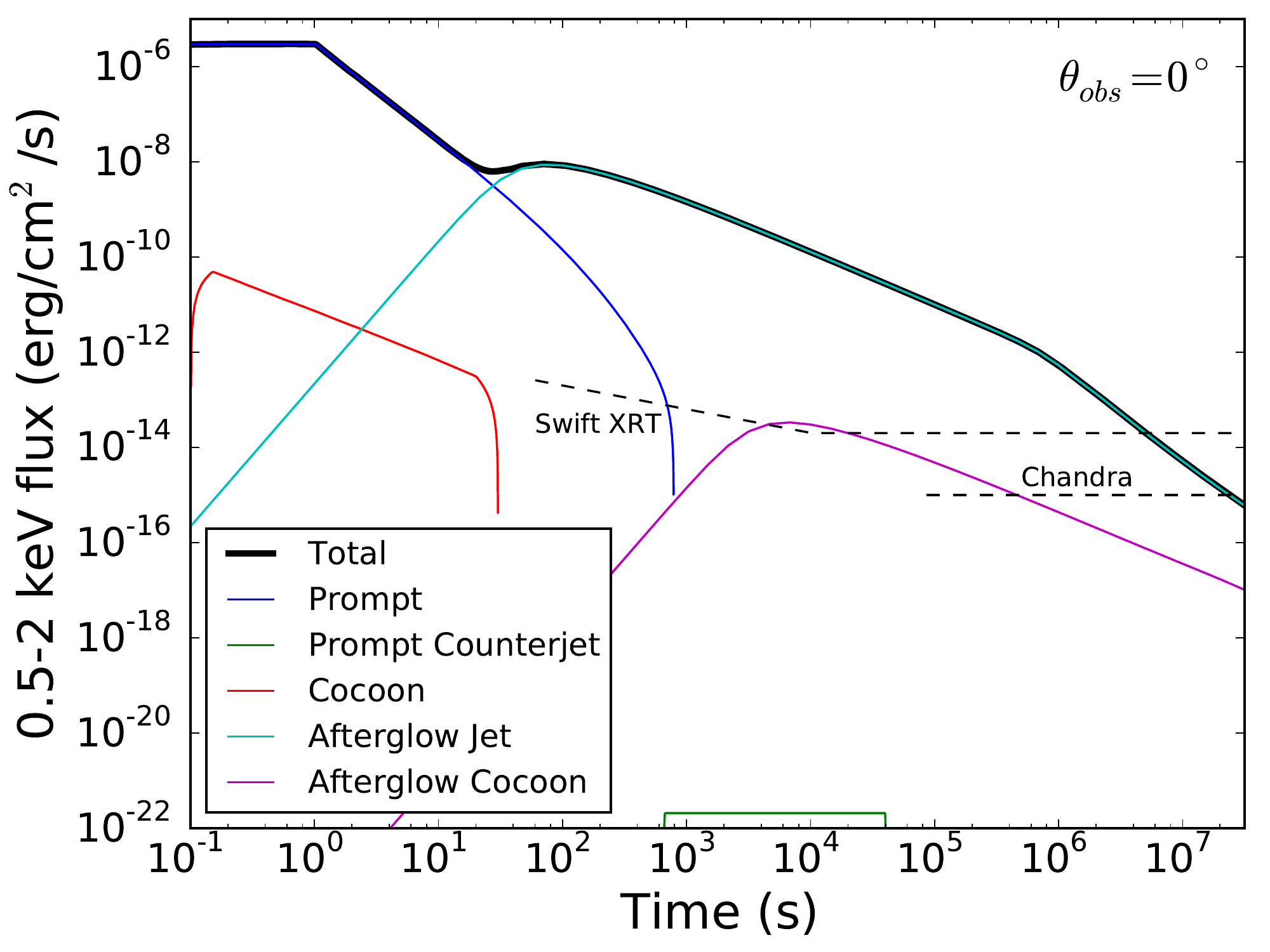}
} \hspace{2truemm}
\parbox{\columnwidth}{
\includegraphics[width=\columnwidth]{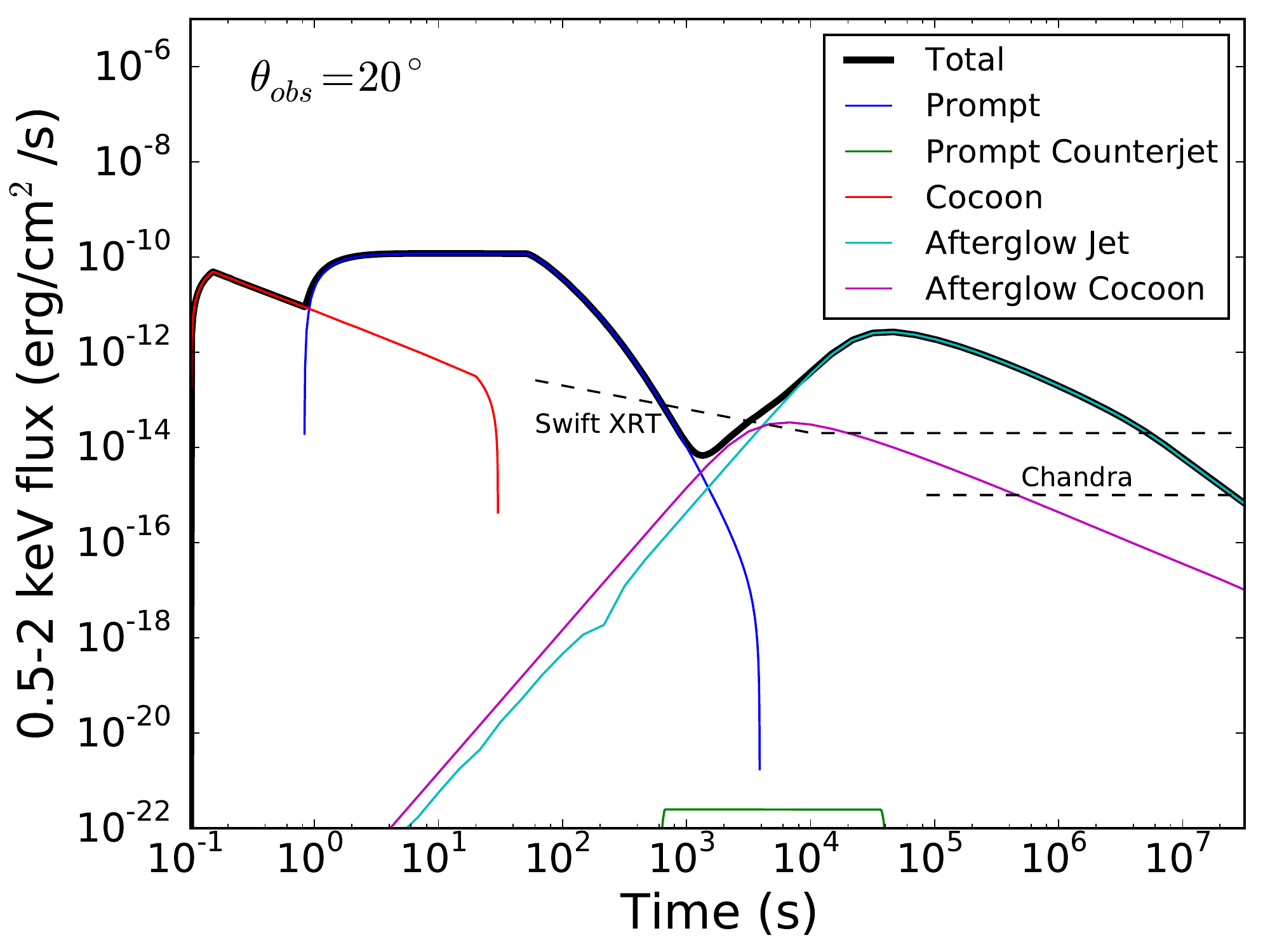}
} 
\parbox{\columnwidth}{
\includegraphics[width=\columnwidth]{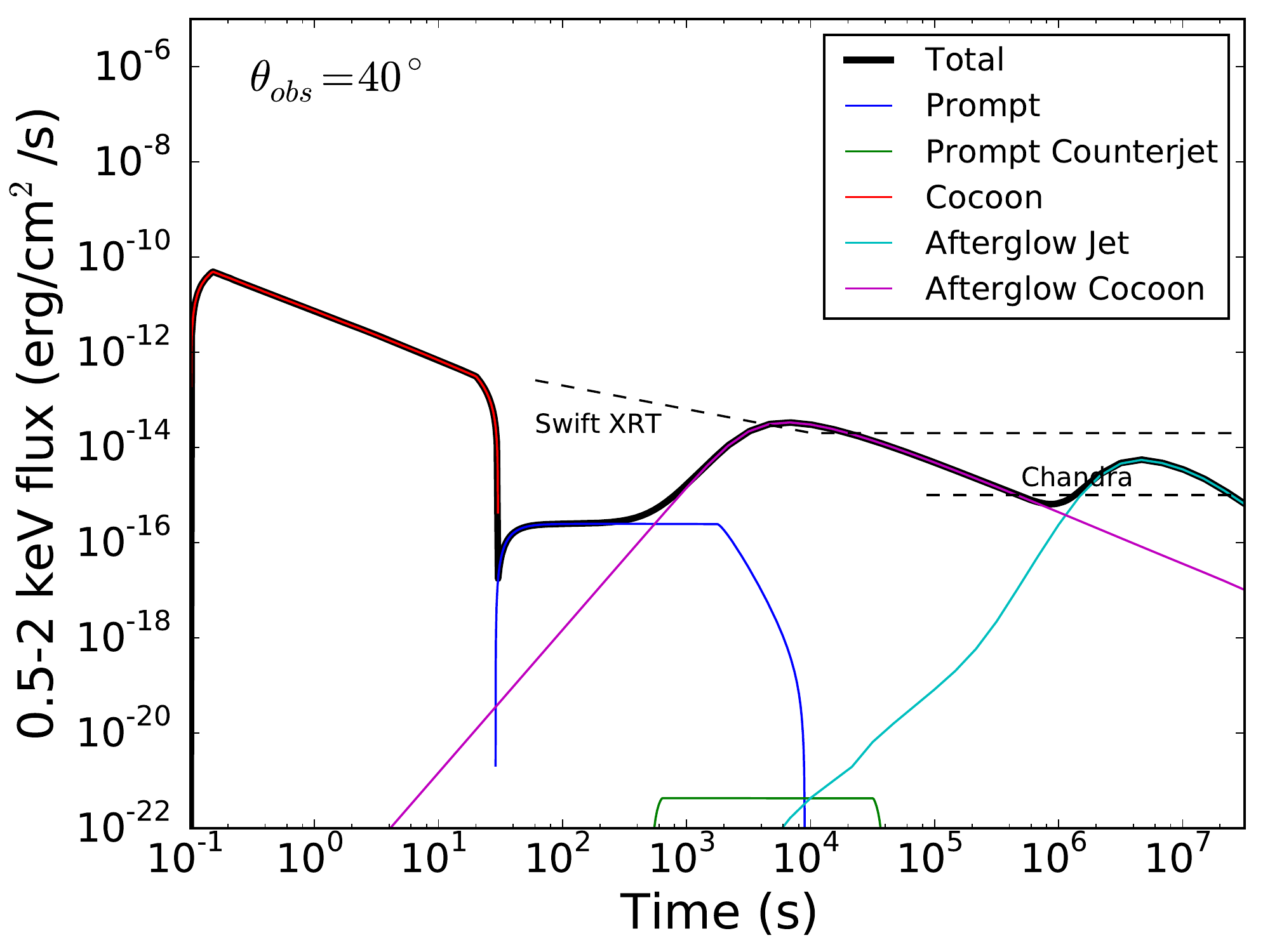}
} \hspace{2truemm}
\parbox{\columnwidth}{
\includegraphics[width=\columnwidth]{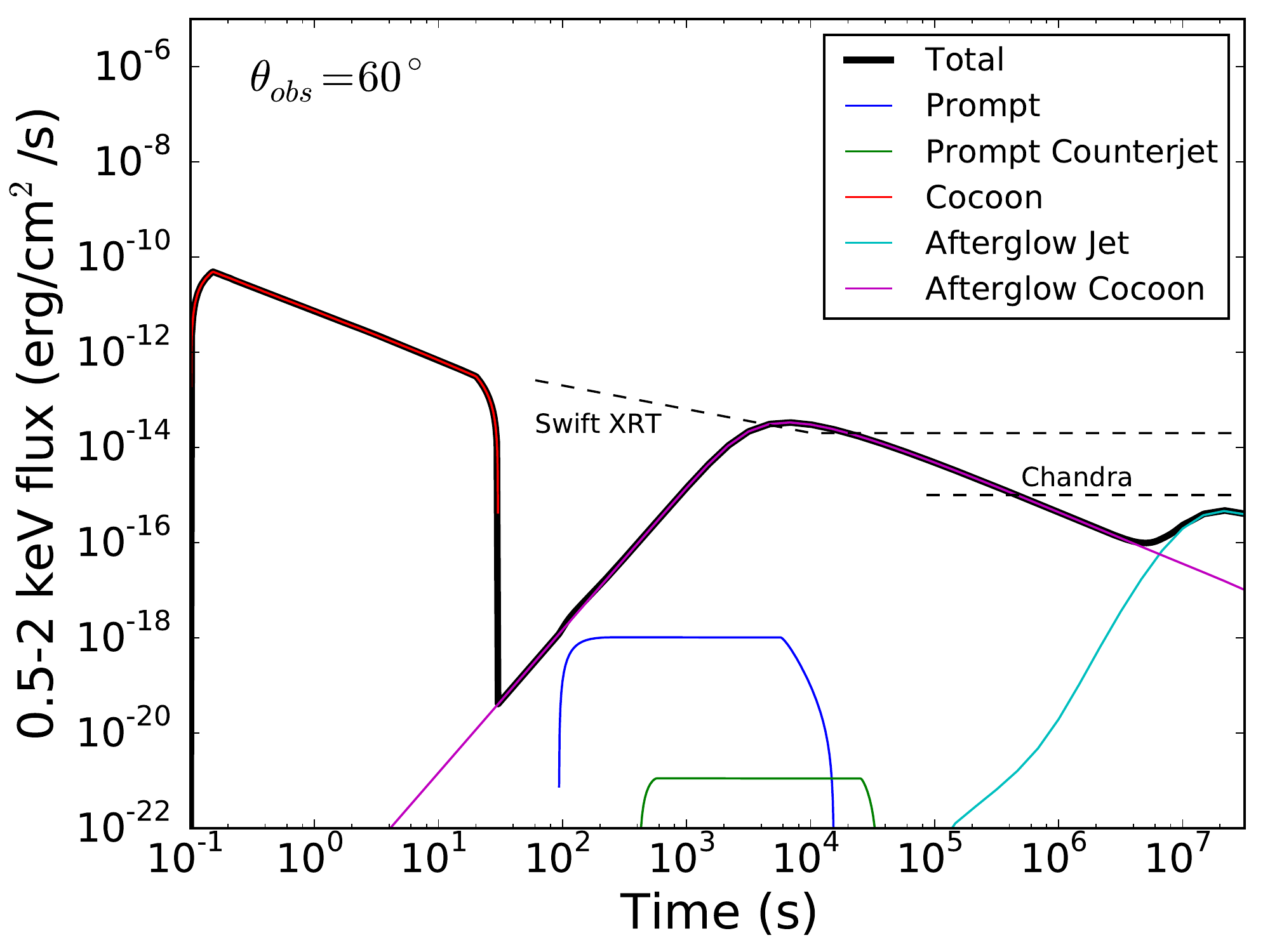}
}
\parbox{\columnwidth}{
\includegraphics[width=\columnwidth]{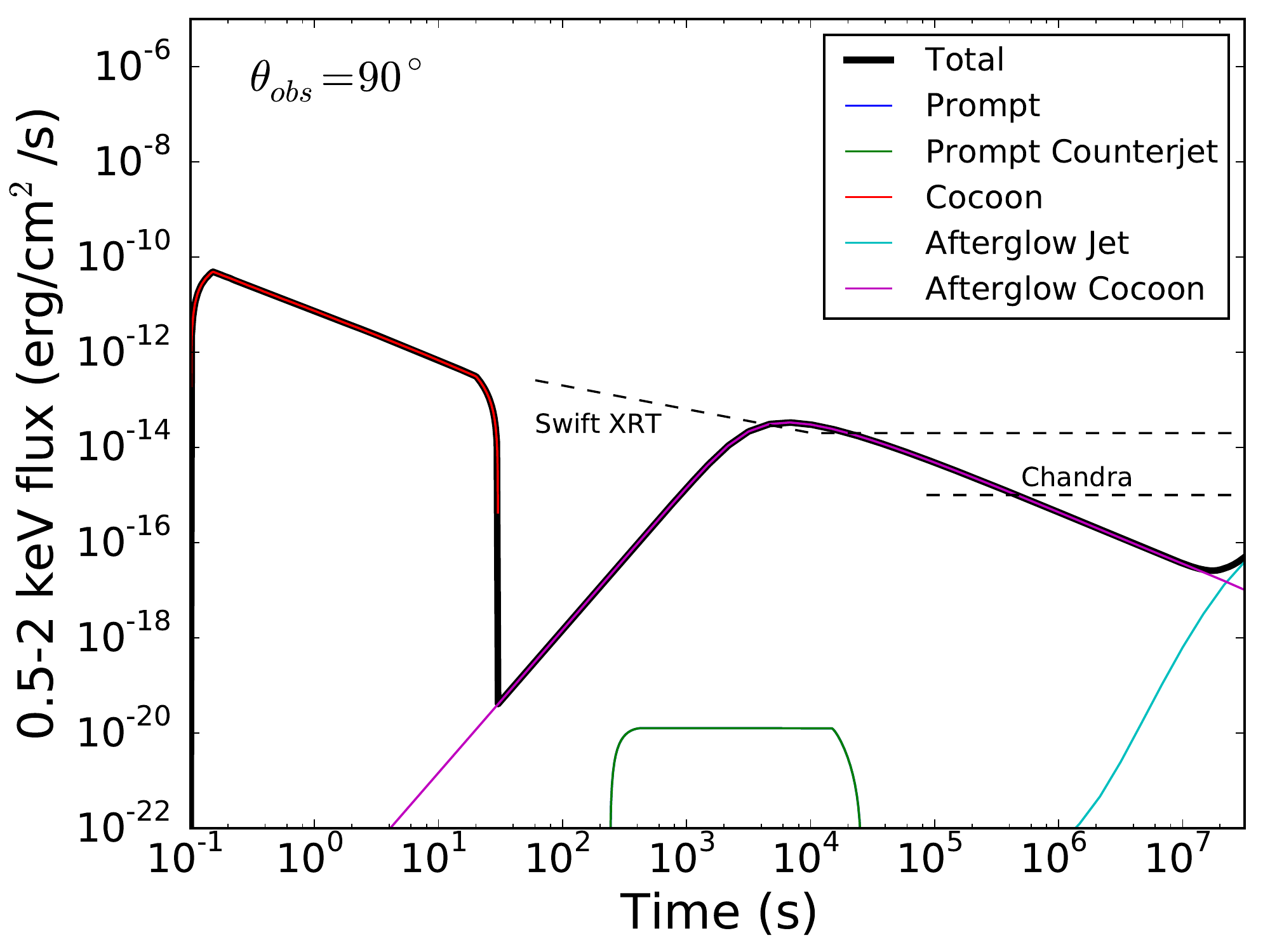}}
\hspace{2truemm}
\parbox{\columnwidth}{\hspace{\columnwidth}
}
\caption{{X-ray light curves from a SGRB at 200 Mpc. Each panel shows
  the various components (thin colored lines) and the total (solid
  black) flux for a unique observer. Observers are located at 0, 20,
  40, 60, and 90$^\circ$ from the jet axis, respectively.}
\label{fig:xray}}
\end{figure*}

The prompt emission light curves in
Figures~\ref{fig:coc_prompt}, ~\ref{fig:xray},~\ref{fig:optical}, and~\ref{fig:radio} were
computed via a Monte Carlo method. Three million emission regions were
generated with random propagation direction within the jet opening
angle. Each of them was turned on at the prescribed distance and given
the Band spectrum described above. The observed activation time of
each emission region was calculated taking into account light
propagation effects, and the observed light curve of each individual
region was calculated by integrating the Band spectrum in the comoving
frequency band corresponding to the observed frequency range. All the
emission regions were then coadded in the final light curves.

\subsection{Cocoon emission}

\subsubsection{Cocoon energy}
We derive the energy that is transferred from the jet into the cocoon
under the assumption that the jet head travels at a constant velocity.
This allows us to write the energy of the cocoon as the luminosity of
two jets, $L_2=E_2/t_{\rm{eng}}$ times the amount of time the jets
spend in the ambient material, $t_{\rm{bo}}$:
$E_{\rm{c}} = L_2t_{\rm{bo}}$. We can express $t_{\rm{bo}}$ as the
distance the jet has to travel, $R_a$, divided by the velocity of the
jet head, $\beta_h$, giving:
\begin{equation}
E_{\rm{c}} = \frac{L_2 R_a}{c\beta_h},
\label{eq:ec}
\end{equation}
which leaves the head velocity, $\beta_h$, as the only quantity to
derive.  We consider two methods for deriving $\beta_h$.  The first
\citep{Morsonyetal2007,Brombergetal2011} assumes that the jet
transversal size is obtained by equating the ram pressure of the jet
material with the pressure of the cocoon. This method should therefore
be appropriate for cocoon-confined jets, those for which the outflow
velocity vector makes a significant angle with the jet-cocoon
discontinuity. They first define the quantity (see also
\citealt{Matzner2003}):
\begin{equation}
\tilde{L} =\frac{L_{2}}{\Sigma_{j} \rho_a c^3},
\end{equation}
the ratio between the energy density of the jet and the energy density
of the surrounding medium, where $L_2$ is the luminosity of the two
jets and $\Sigma_{j}=R_a^2\Omega_2$ is the cross-sectional area of the
jet at breakout. By making cylindrical approximations for the jet, they show it is
approximately

\begin{equation}
	\tilde{L} \simeq \left(\frac{L_2\pi^2}{\rho_a t_\text{bo}^2 \Omega_2^2 c^5}\right)^{2/5}.
\end{equation}
Then, following \cite{Matzner2003}, we find
\begin{equation}
\beta_h = \frac{\beta_j}{1+\tilde{L}^{-1/2}},
\end{equation}
For the values considered in this paper (see Table~\ref{tab:symbols}),
$\tilde{L}\ll1$ and therefore
\begin{align}
	\beta_h &\simeq \tilde{L}^{1/2}.
\label{eq:betahV1}
\end{align}
Therefore, with $t_\text{bo} = R_{\rm{a}}/c\beta_h$, the energy in the cocoon is
\begin{equation}
	E_{\rm{c}} = \left(L_2^2R_a^5 \rho_a \Omega_2^2\pi^{-2}\right)^{1/3}.
\label{eq:bromb-ec}
\end{equation}

Using the values presented in Table~\ref{tab:symbols}, this gives an
energy of $E_{\rm{c}} = 4.2\times 10^{48}\text{ erg}$.

An alternative derivation of $\beta_h$ can be obtained by balancing
the jet thermal pressure and the cocoon pressure to compute the jet
transversal size \citep{LazzatiBegelman2005,Lazzatietal2012}. This
approximation holds for mildly confined jets, for which the ram
pressure is negligible due to the fact that the relativistic outflow
velocity is nearly parallel to the jet-cocoon discontinuity. They
find:
\begin{equation}
\beta_h = \left(\frac{2 L_2^3}{\pi c^9 r_0^4 R_a^2 \Omega_2^2
    \rho_a^3}\right)^{1/7},
\label{eq:betahV2}
\end{equation}
which gives a cocoon energy:
\begin{equation}\label{eq:lazec}
E_{\rm{c}} = \left(\frac{L_2^4 R_a^9 r_0^4 \Omega_2^2 \rho_a^3 c^2 \pi}{2}\right)^{1/7},
\end{equation}
which gives an energy of $E_{\rm{c}} \simeq 7\times 10^{48}$~erg for
the fiducial values reported in Table~\ref{tab:symbols}.

The two results for the cocoon energy are fairly similar. Not
surprisingly, the second value gives a higher cocoon energy, since the
jets are only mildly confined by the relatively low density of the
merger ejecta.  In the following we adopt a fiducial value
$E_{\rm{c}} =10^{49}$~erg for the cocoon energy. We note that the
cocoon energy depends less than linearly on all jet and ejecta
properties with the exception of the radius of the polluted
region. Should the ejecta be distributed in a larger region than the
one assumed here, the cocoon energy would be significantly higher. To
test the dependency of the detectability of the cocoon on the
uncertain energetics, we subsequently test values of the cocoon energy
ten times lower and higher than our fiducial value.

\begin{figure*}
\parbox{\columnwidth}{
\includegraphics[width=\columnwidth]{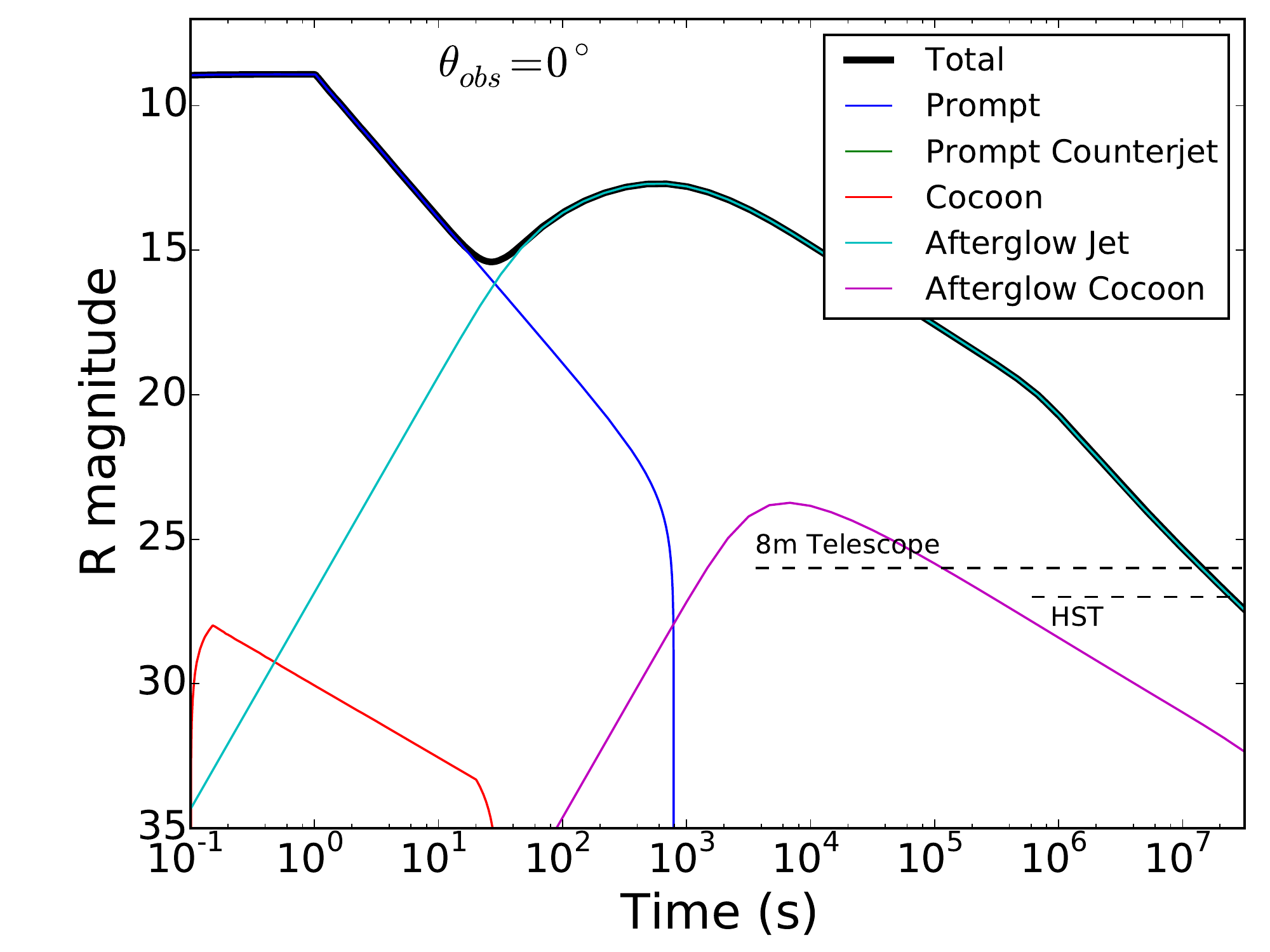}
} \hspace{2truemm}
\parbox{\columnwidth}{
\includegraphics[width=\columnwidth]{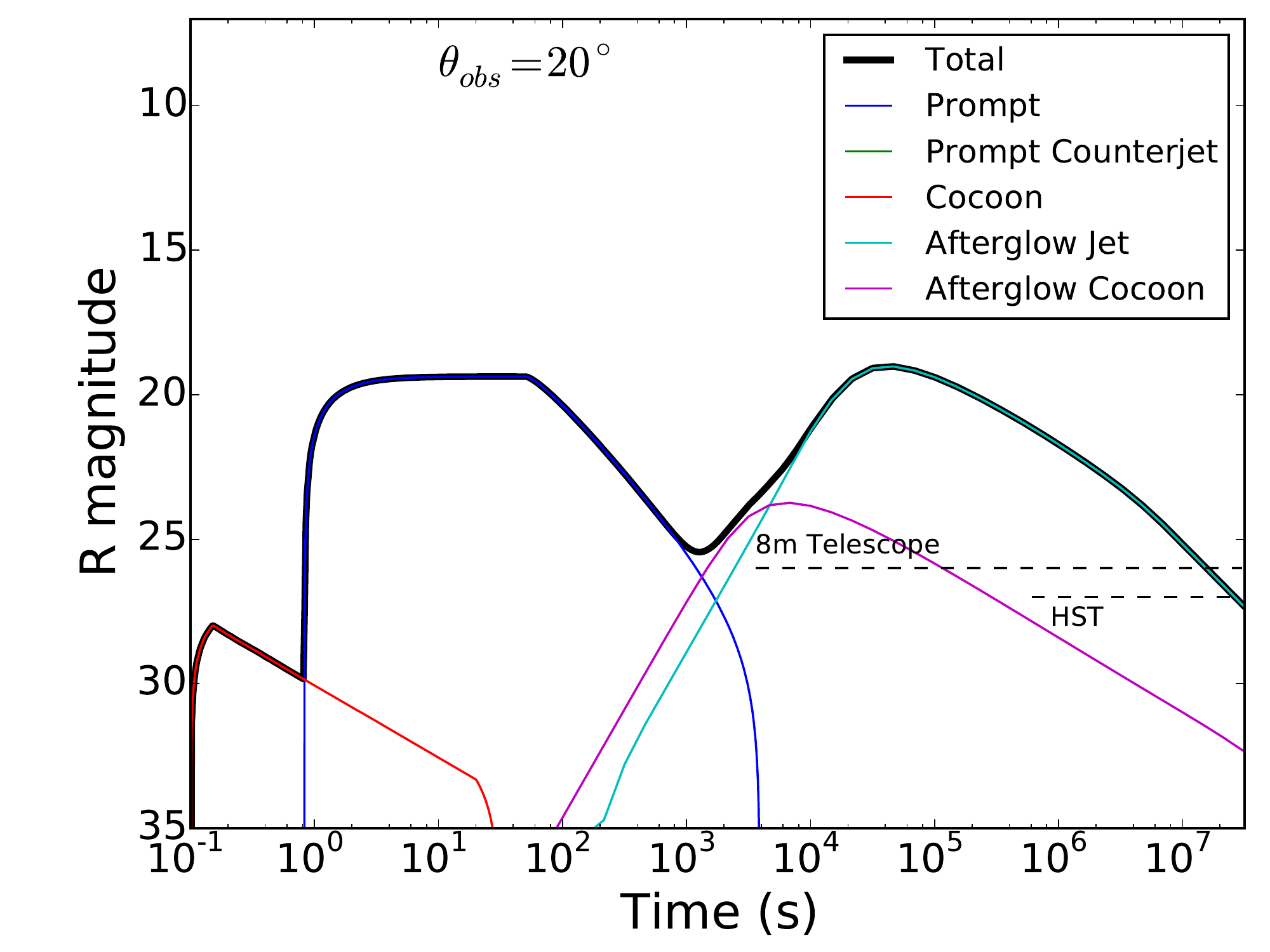}
} 
\parbox{\columnwidth}{
\includegraphics[width=\columnwidth]{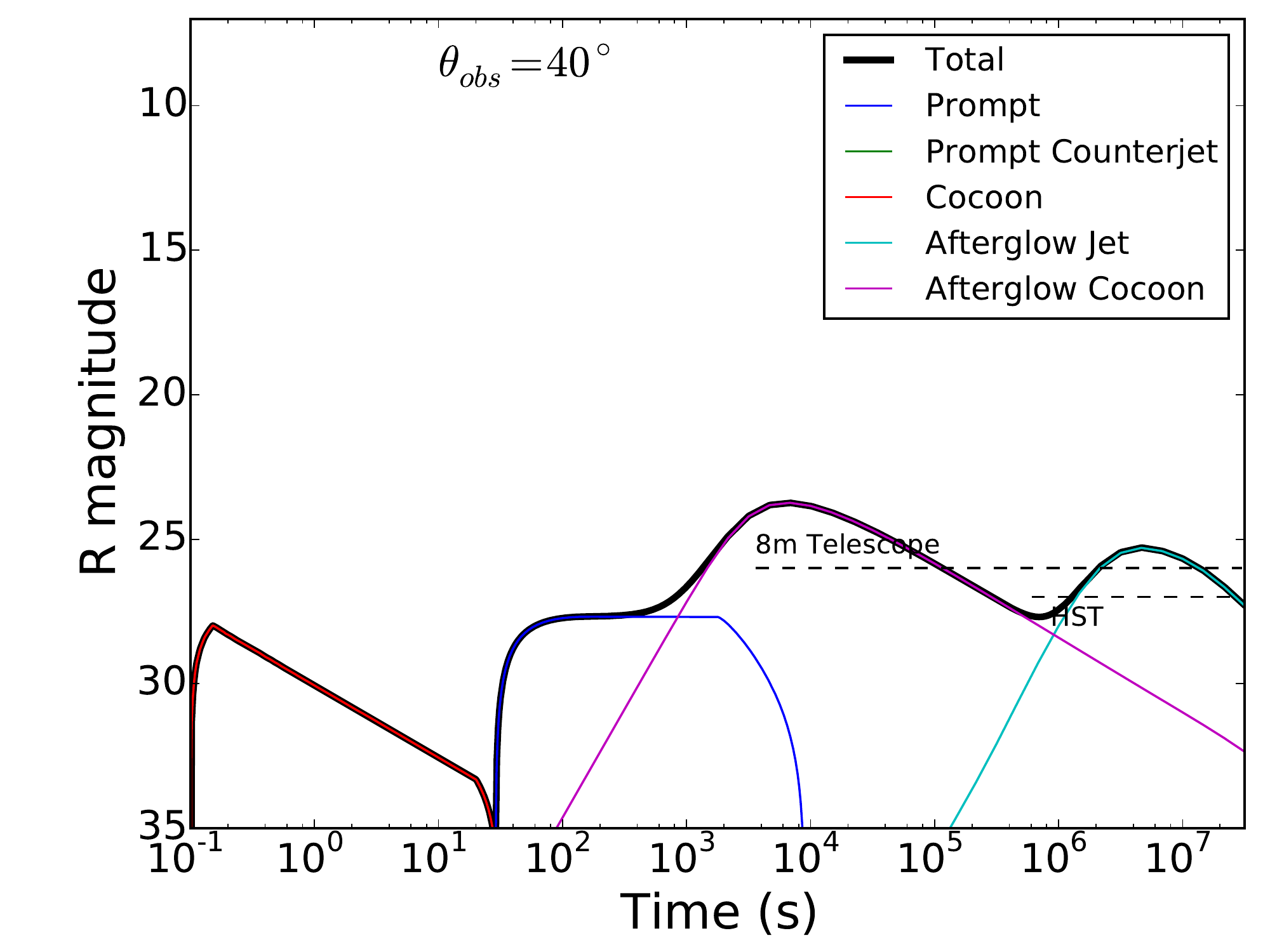}
} \hspace{2truemm}
\parbox{\columnwidth}{
\includegraphics[width=\columnwidth]{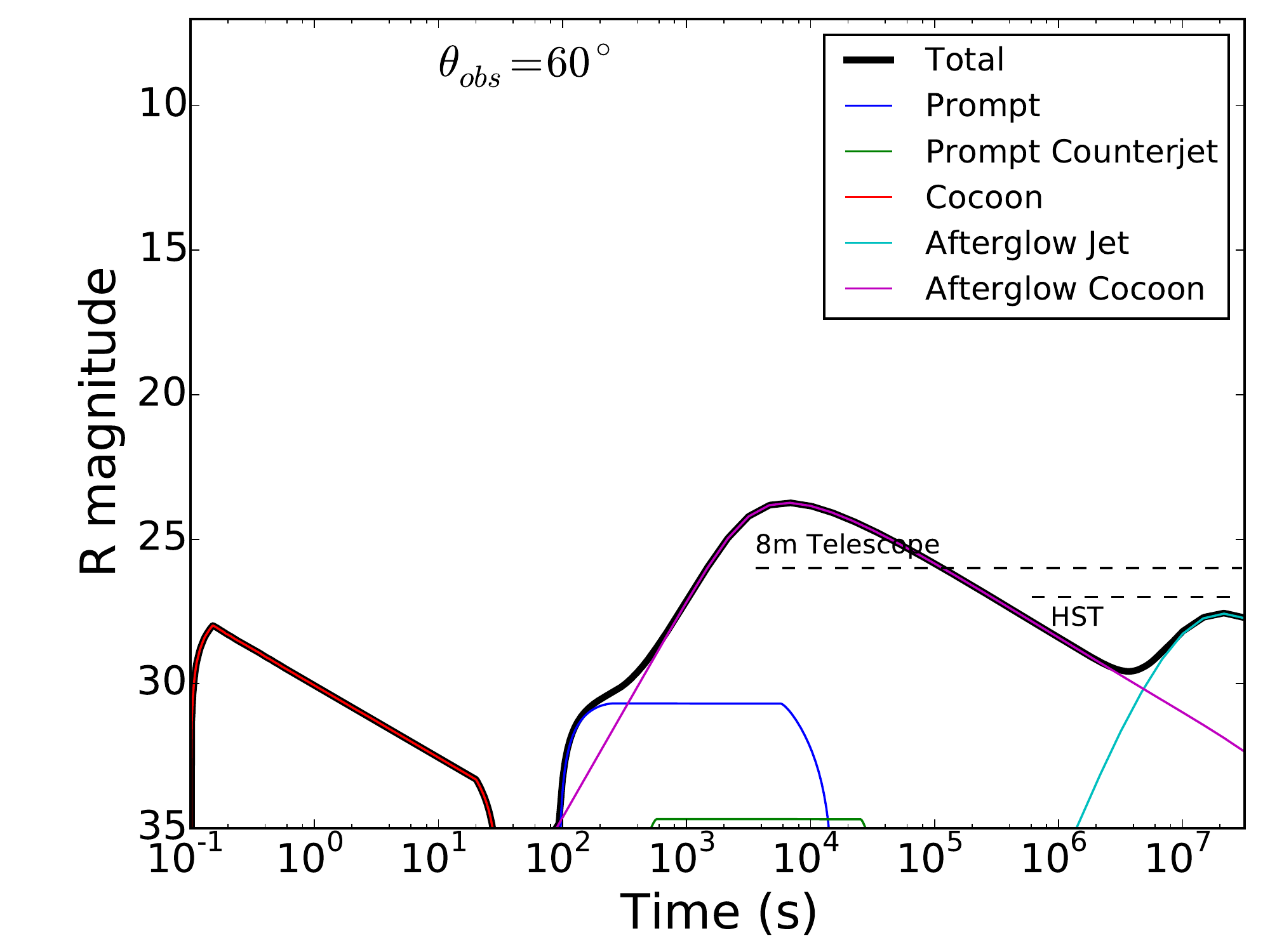}
}
\parbox{\columnwidth}{
\includegraphics[width=\columnwidth]{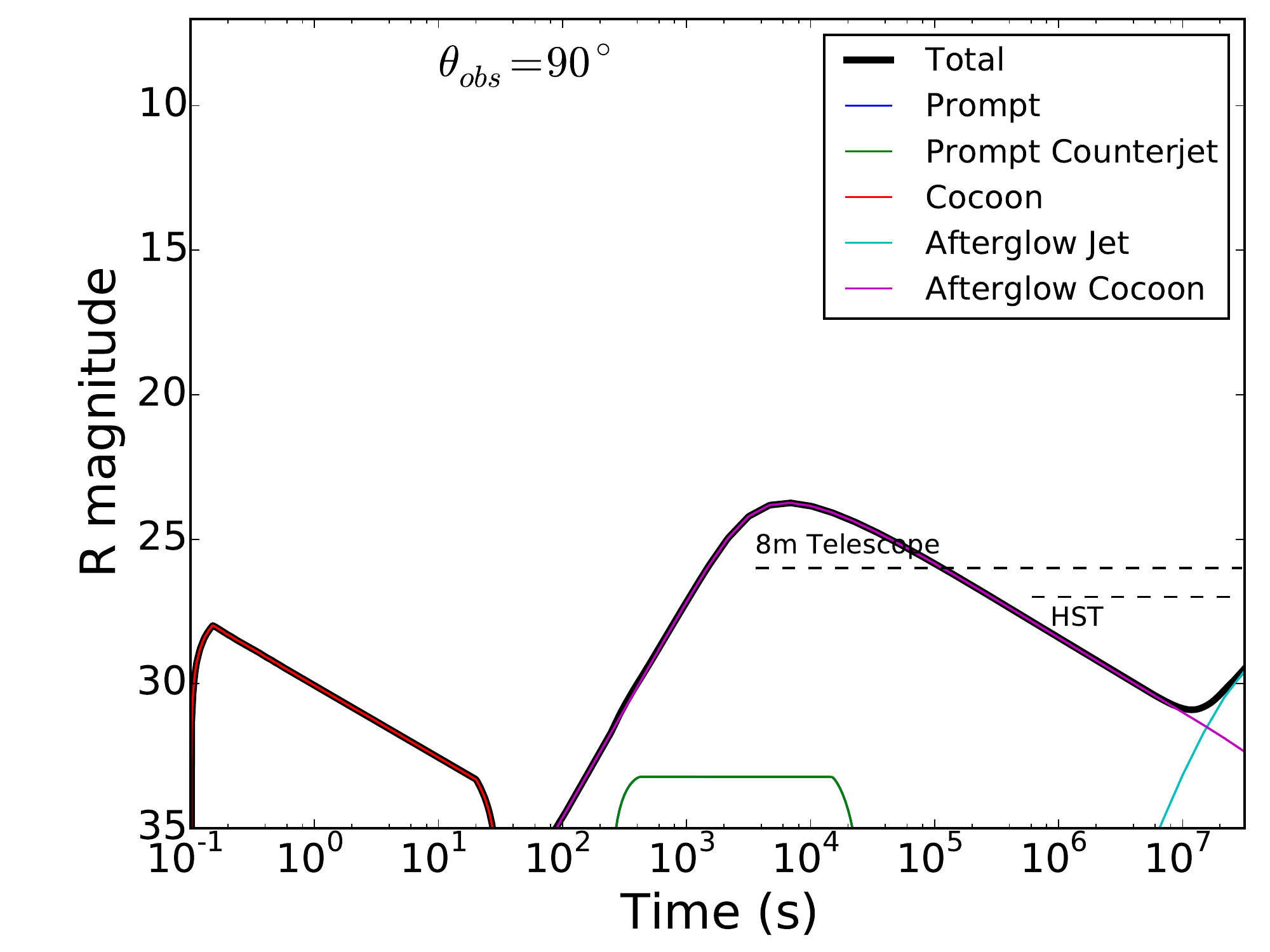}}
\hspace{2truemm}
\parbox{\columnwidth}{\hspace{\columnwidth}
}
\caption{{Same as Figure~\ref{fig:xray} but showing R-band magnitudes.}
\label{fig:optical}}
\end{figure*}

\subsubsection{Cocoon dynamics and radiation}

\begin{figure*}
\parbox{\columnwidth}{
\includegraphics[width=\columnwidth]{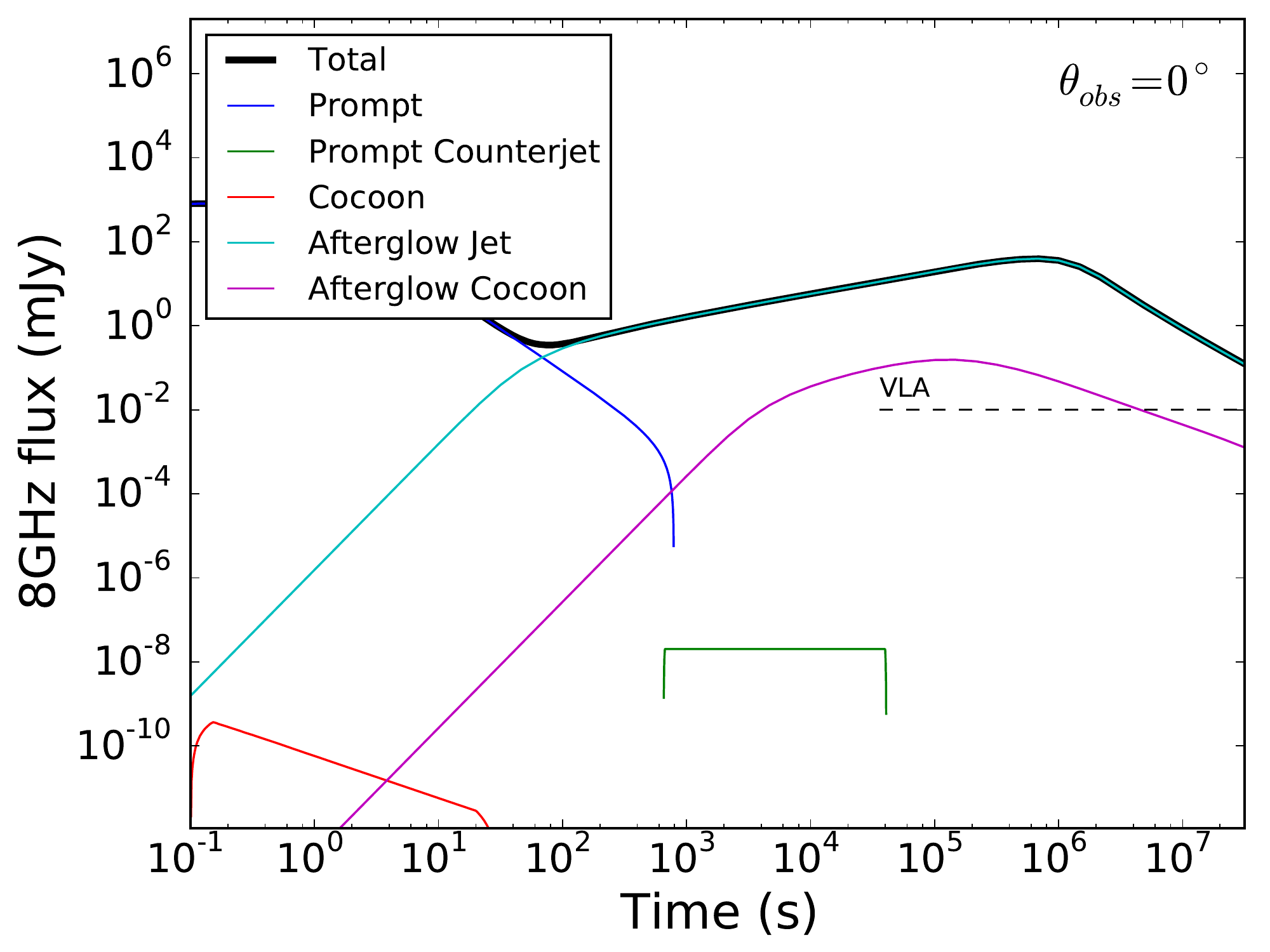}
} \hspace{2truemm}
\parbox{\columnwidth}{
\includegraphics[width=\columnwidth]{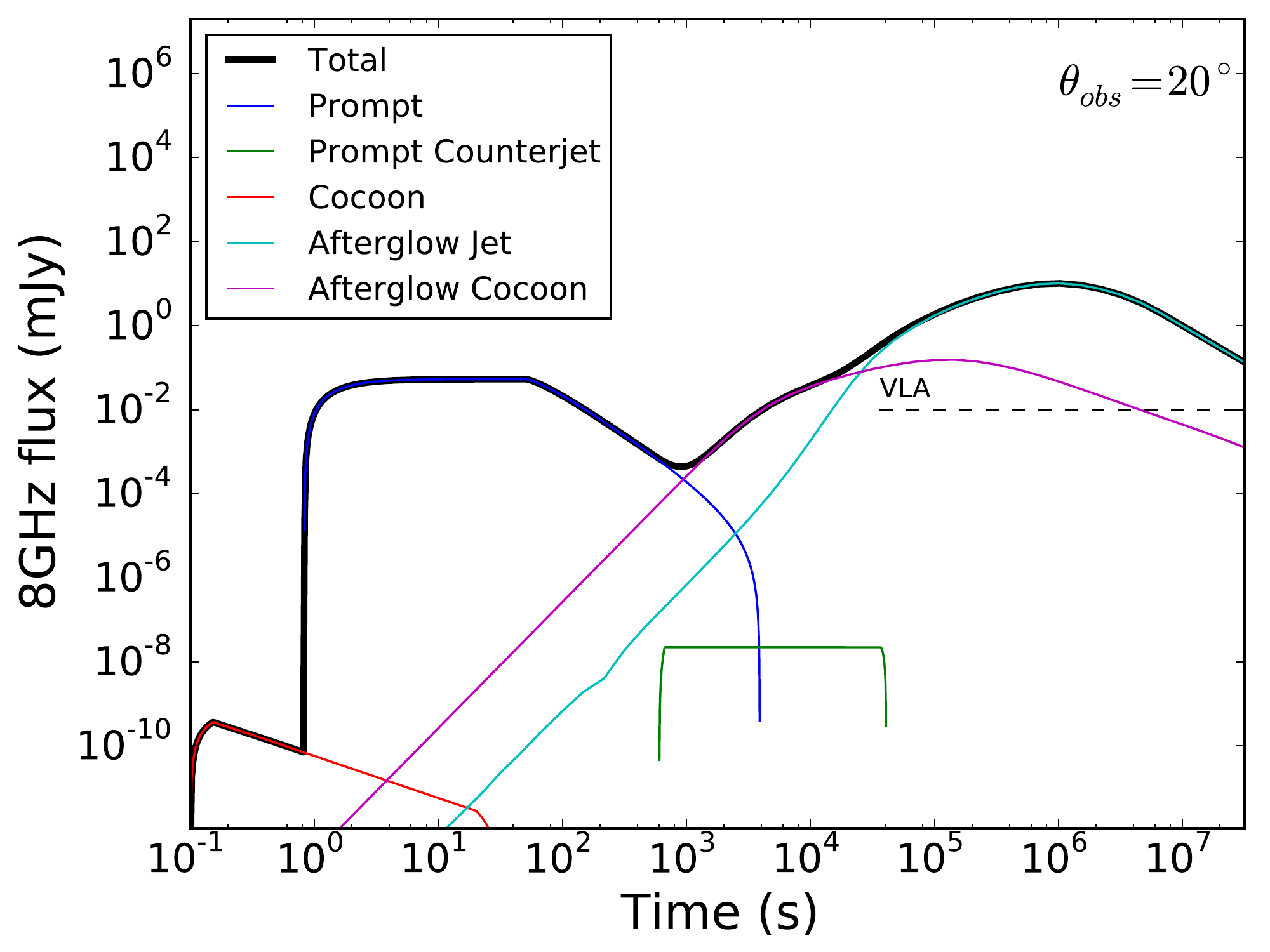}
} 
\parbox{\columnwidth}{
\includegraphics[width=\columnwidth]{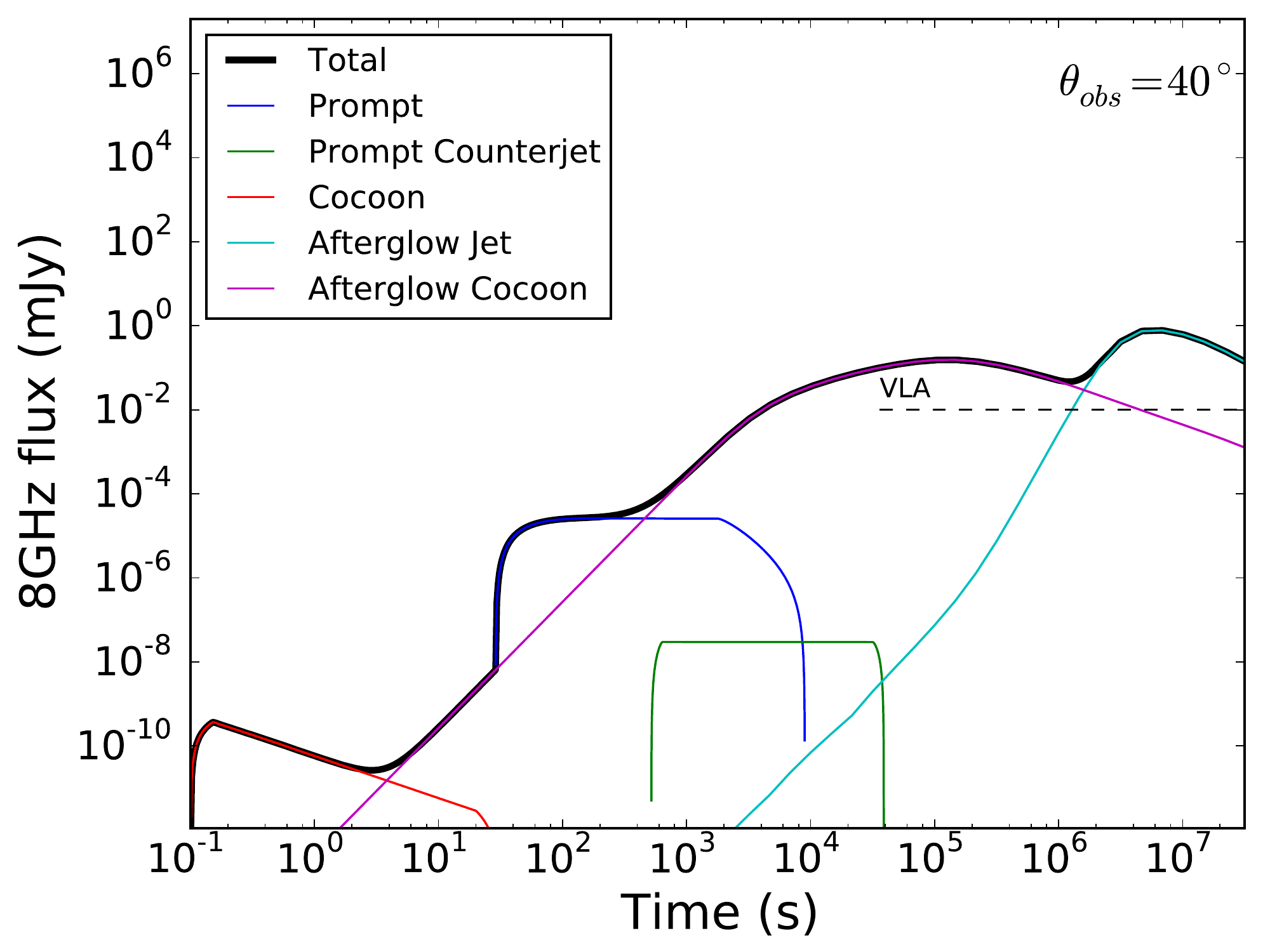}
} \hspace{2truemm}
\parbox{\columnwidth}{
\includegraphics[width=\columnwidth]{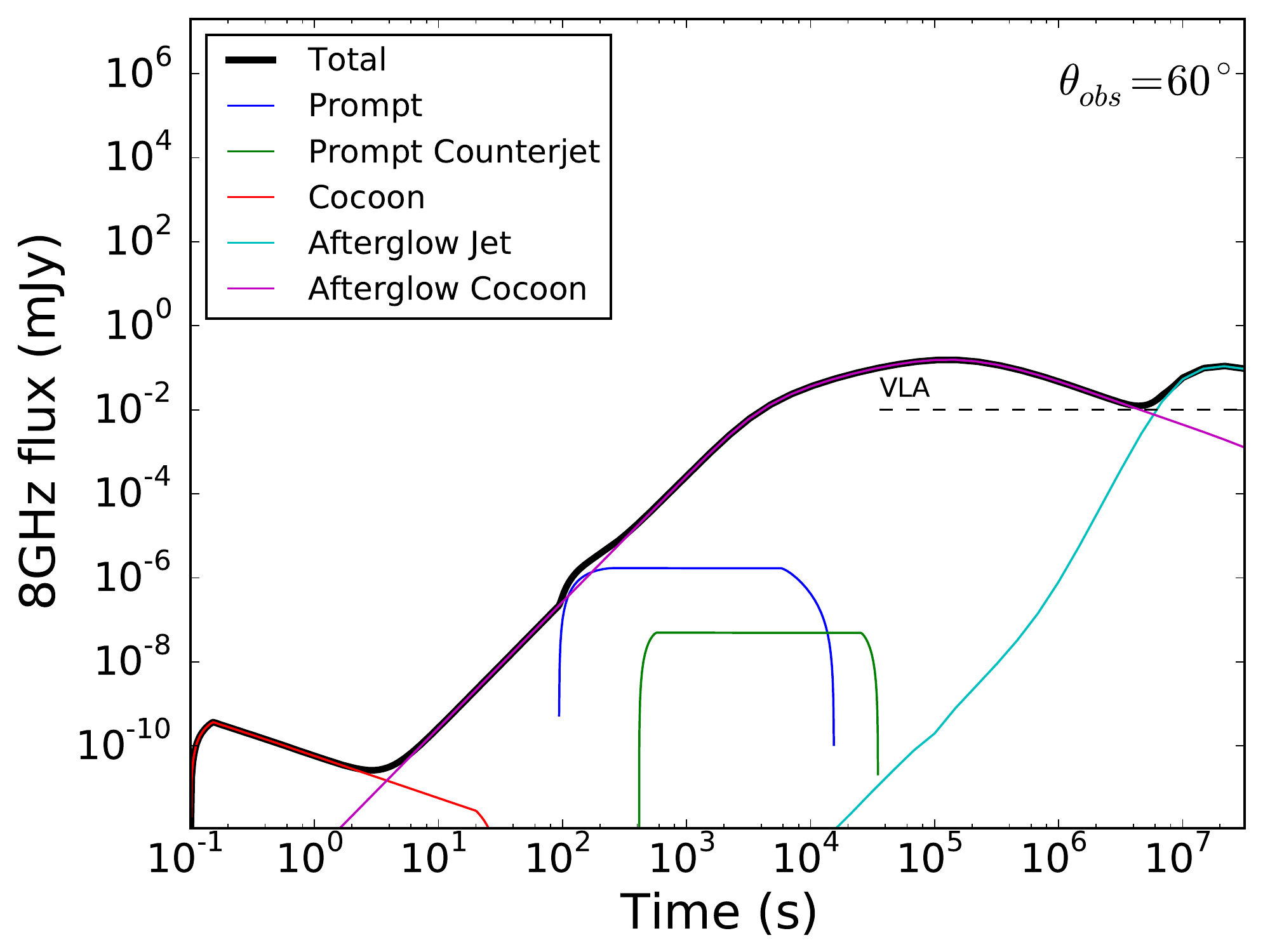}
}
\parbox{\columnwidth}{
\includegraphics[width=\columnwidth]{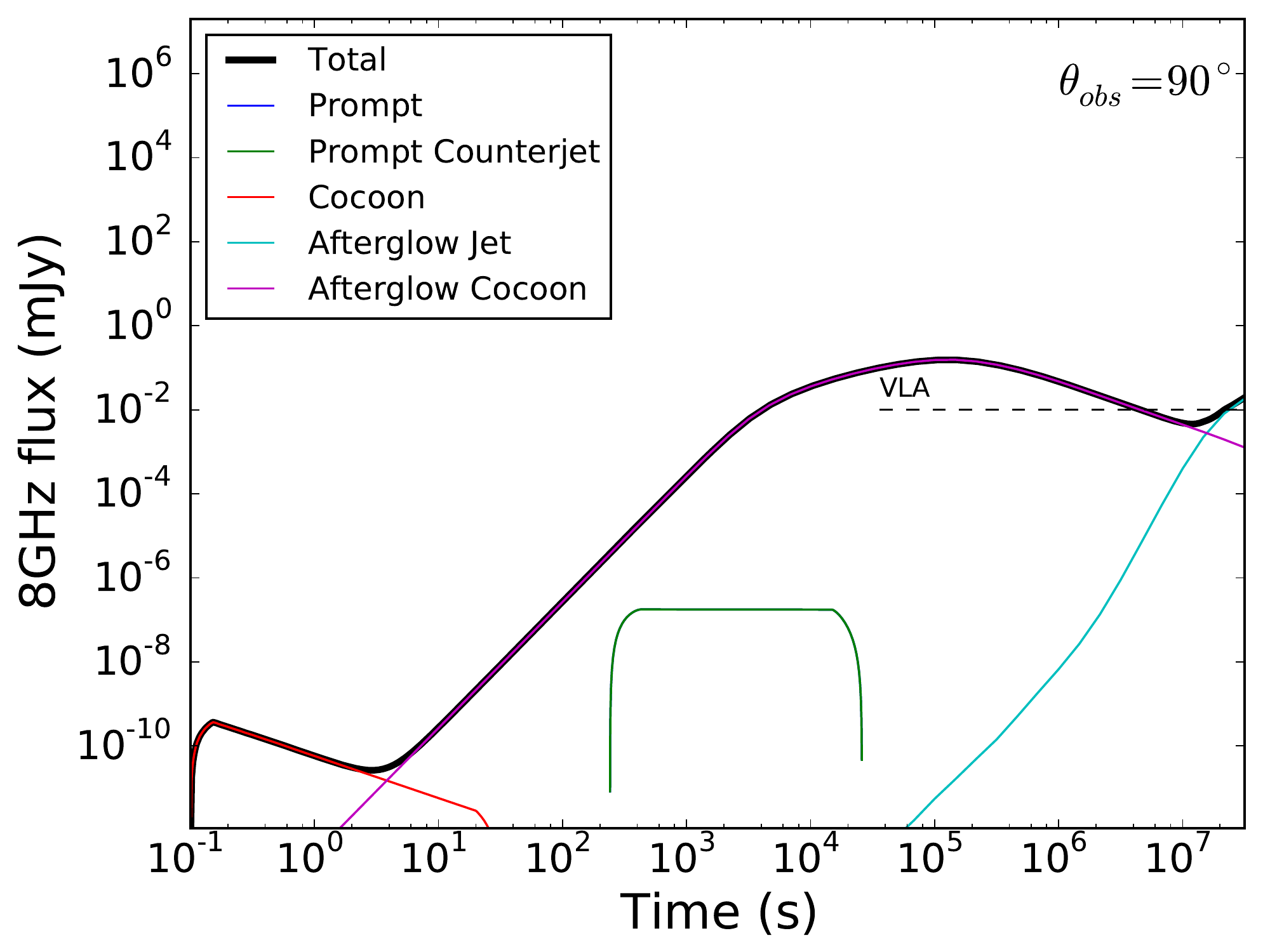}}
\hspace{2truemm}
\parbox{\columnwidth}{\hspace{\columnwidth}
}
\caption{{Same as Figure~\ref{fig:xray} but showing 8~GHz flux densities.}
\label{fig:radio}}
\end{figure*} 

Let us now consider the subsequent evolution of the cocoon. At
breakout, the cocoon is hot and high-pressured with no associated bulk
motion. Upon release, it accelerates quasi-isotropically
\citep{RamirezRuizetal2002,NakarPiran2017}. The light curve produced
by the cocoon depends critically on the contamination from the ambient
material. Numerical simulations of long GRB jets show that the cocoon
is polluted by the progenitor material in such a way that it has an
asymptotic Lorentz factor $\Gamma_{\infty,{\rm{c}}}\sim10$
\citep{Lazzatietal2010}. We therefore adopt a value
$\Gamma_{\infty,{\rm{c}}}=10$ in the remainder of this study, but we
warn the reader that dedicated numerical simulations should be
performed to pin down the actual composition of SGRB cocoons (see,
e.g., \cite{Gottliebetal2017}).

Initially, the cocoon fireball accelerates self similarly,
$\Gamma\propto r$, until the saturation radius (e.g.,
\citealt{CavalloRees1978,ChhotrayLazzati2017}):
\begin{equation}
R_{\rm{sat,c}}=\Gamma_{\infty,{\rm{c}}}R_a
\end{equation}

Beyond the saturation radius the cocoon fireball coasts at constant
Lorentz factor, releasing the advected radiation at the photospheric
radius \citep{MeszarosRees2000}:
\begin{equation}
R_{\rm{ph,c}}=\left(\frac{E_c\sigma_T}{8\pi m_p
    \Gamma_{\infty,{\rm{c}}}^3 c^2}\right)^\frac12
\label{eq:rph}
\end{equation}
which is, for the parameter choices of Table~\ref{tab:symbols}, beyond
the saturation radius $R_{\rm{sat,c}}$. The observed cocoon
temperature is constant during the acceleration phase, and scales as
$r^{-2/3}$ in the coasting phase, yielding a photospheric temperature:
\begin{equation}
T_{\rm{ph,c}}=T_{0,{\rm{c}}}\left(\frac{R_{\rm{ph,c}}}{R_{\rm{sat,c}}}\right)^{-\frac23}
=\left(\frac{E_c}{aV_c}\right)^\frac14 
\left(\frac{E_c\sigma_T}{8\pi m_p
    R_{\rm{a}}^2\Gamma_{\infty,{\rm{c}}}^5 c^2}\right)^{-\frac13}
\label{eq:t0}
\end{equation}
where $T_{0,c}=(E_{\rm{c}}/aV_{\rm{c}})^{1/4}$ is the initial temperature of the cocoon, $a$
is the radiation constant and $V_{\rm{c}}$ is the cocoon volume at
breakout time. Following Lazzati \& Begelman (2005) we find:
\begin{equation}
V_{\rm{c}}=\frac{2}{3\beta_{\rm{h}}c}\sqrt{\frac{\pi E_{\rm{c}}R_{\rm{a}}^3}{\rho_{\rm{a}}}}
\end{equation}
yielding an observed cocoon photospheric temperature:
\begin{equation}
T_{\rm{ph,c}}=\left(\frac{24}{a}\right)^\frac14
\pi^\frac{5}{24}c^\frac{11}{12}\left(\frac{m_p}{\sigma_T}\right)^\frac13
\frac{\beta_{\rm{h}}^\frac14\,\rho_{\rm{a}}^\frac18\,
\Gamma_{\infty,{\rm{c}}}^\frac53\,R_{\rm{a}}^\frac{7}{24}}{E_{\rm{c}}^\frac{5}{24}}
\end{equation}
Adopting the fiducial SGRB values from Table~\ref{tab:symbols} and the
jet head velocity from either Equation~\ref{eq:betahV1}
or~\ref{eq:betahV2} we find $T_{\rm{ph,c}}\simeq10$~keV, the value
that we adopt for the light curves calculations.

Assuming that the radiated spectrum is a black body\footnote{In the
  absence of internal dissipation and magnetic fields, the spectrum is
  expected to be a broadened black body
  \citep{Goodman1986,Lazzati2016,DeColleetal2017}.}  we obtain a
luminosity:
\begin{equation}
L_{\rm{c}}=\sigma\pi\left(\frac{R_{\rm{ph,c}}}{\Gamma_{\infty,{\rm{c}}}}\right)^2T_{\rm{ph,c}}^4
=
\frac{3 m_p^\frac13 \pi^\frac56 \sigma c^\frac53}{\sigma_T^\frac{1}{3} \,
  a} E_{\rm{c}}^\frac16
\Gamma_{\infty,{\rm{c}}}^\frac53\beta_{\rm{h}}\rho_{\rm{a}}^\frac12 R_{\rm{a}}^\frac76
\end{equation}
where we have used Equations~\ref{eq:rph} and~\ref{eq:t0} for the
final result. For the typical SGRB values adopted above, we have:
\begin{equation}
L_{\rm{c}}=4\times10^{49}
\left(\frac{E_c}{10^{49}}\right)^\frac23
\left(\frac{\Gamma_{\infty,{\rm{c}}}}{10}\right)^\frac53 
\left(\frac{R_{\rm{a}}}{10^8}\right)^{-\frac13} 
\end{equation}

The emission would last for the longest time between the diffusion
time in the cocoon shell $\delta t_{\rm{diff}}\sim R_{\rm{a}}/c$ and
the angular time scale
$\delta t_{\rm{ang}}=R_{\rm{ph,c}}/(c
\Gamma_{\infty,{\rm{c}}}^2)$.
For $\Gamma_{\infty,{\rm{c}}}=10$ the latter dominates and the
cocoon thermal pulse would last $\sim0.3$~s. Note that multiplying the
photospheric luminosity times the pulse duration we obtain a
photospheric radiative energy that is comparable (within a factor of
order unity) to the cocoon energy, as expected.

The light curves of the cocoon photospheric emission shown in
Figures~\ref{fig:coc_prompt}, ~\ref{fig:xray},~\ref{fig:optical},
and~\ref{fig:radio} were computed via the same Monte Carlo method
discussed for the prompt emission. The comoving spectrum, however, was
assumed to be thermal with temperature $T_{\rm{ph,c}}^\prime=1$~keV.

\subsection{Afterglow}

The afterglow emission of the jet and cocoon components are calculated
using the semi-analytic Trans-Relativistic Afterglow Code (TRAC),
identical to that used in \cite{Morsonyetal2016} (full description
will be published in Morsony et al. in preparation).  TRAC is able to model
the emission of a relativistic fireball with an arbitrary energy
distribution, as seen by an observer at any angle relative to the jet
axis.  We assume all afterglow emission is produced by synchrotron
radiation, including synchrotron self-absorption and local synchrotron
cooling.  For all models presented here, synchrotron radiation is
parameterized by $\epsilon_e=0.1$, the fraction of energy in
electrons, $\epsilon_B=0.01$, the fraction of energy in the magnetic
field, and $p=2.5$, the spectral index of the electron energy
distribution.

We model the jet and cocoon as expanding into a constant density
external medium (ISM) with number density of
$n_{\rm{ISM}}=10^{-1}$~cm$^{-3}$.  The cocoon component is modeled as
a spherical explosion with kinetic energy of $E_{\rm{c}}=10^{49}$~erg
and initial Lorentz factor of $\Gamma_{\infty,c}=10$.  The jet is
modeled as a top-hat jet with isotropic-equivalent kinetic energy of
$E_{\rm{iso}}=2.5 \times 10^{51}$~erg (half of the initial energy,
since 50 per cent of it was released as gamma-ray radiation in the
prompt emission) and initial Lorentz factor $\Gamma_{\infty}=100$
within an half opening angle $\theta_{j}=16\degr$ and no material
outside the jet.

\section{Results}

Figures~\ref{fig:coc_prompt}, ~\ref{fig:xray},~\ref{fig:optical},
and~\ref{fig:radio} summarize the results of our calculations.  Figure
~\ref{fig:coc_prompt} shows how the prompt emission of an off-axis
SGRB would be seen by wide field X-ray and $\gamma$-ray monitors on
board Swift and Fermi.

The three other figures show how a burst with the properties listed in
Table~\ref{tab:symbols} and located at 200 Mpc from Earth would be
observed in X-rays, optical, and radio bands, respectively. Each
figure has five panels, each panel showing the observations that would
be performed by an observer located along a particular line of
sight. From left to right and top to bottom, observers at 0, 20, 40,
60, and 90 degrees from the jet axis are shown. Along with the
predicted fluxes and flux densities, we show the detection limit of
instruments that could be used to search for the electromagnetic
counterpart of the GW event. In X-ray, we consider Swift XRT and
Chandra. For late XRT observations we assume a detection limit of
$2\times10^{-14}$ erg~cm$^{-2}$~s$^{-1}$ (for a 10~ks exposure). For
early XRT observations, we assume a detection limit that scales with
the square root of the exposure time, assuming that the signal to
noise is background dominated. A constant detection limit of
$10^{-15}$ erg~cm$^{-2}$~s$^{-1}$ is adopted for Chandra,
corresponding to a $\sim50$~ks exposure. We also assumed that Swift
can repoint XRT to the burst location within one minute, while it
takes Chandra one day to repoint. In the optical we show R-band
imaging detection limits for an 8 meter class telescope and for HST,
R=26 and 27, respectively. Finally, in the radio, we report a VLA
detection limit of 10~$\mu$Jy, assuming a 10 hours reaction time.

\begin{figure*}
\includegraphics[width=\textwidth]{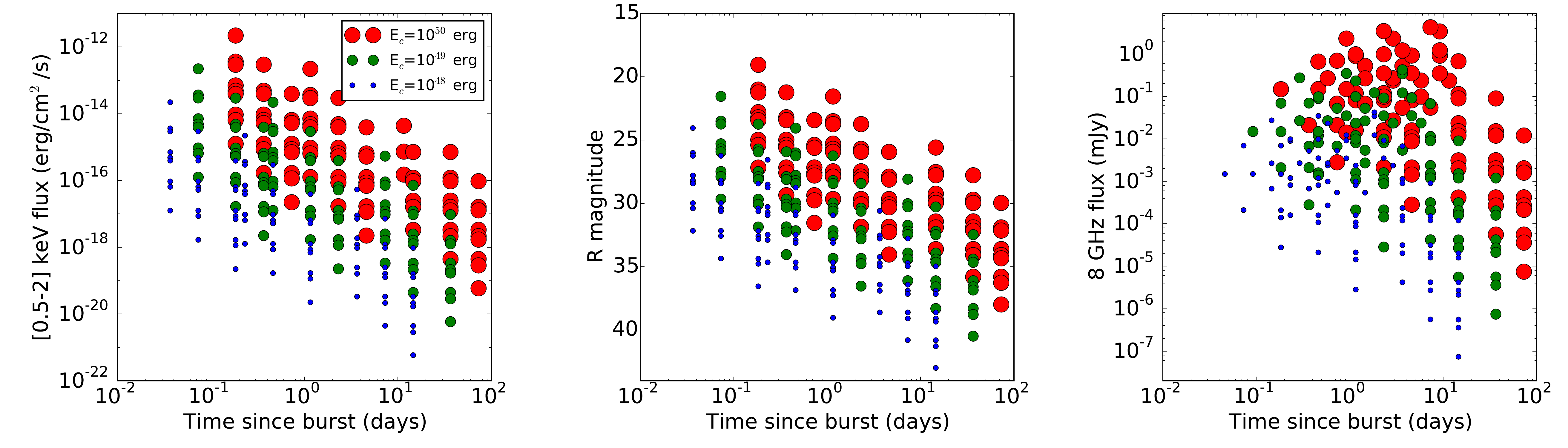}
\caption{{Peak emission fluxes vs. peak times for cocoon afterglows
  from a range of cocoon and external shock properties (see text). The
  unknown cocoon physics and properties of the external shock cause a
  significant uncertainty on the detectability of the EM transient
  from off-axis observers.}
\label{fig:new}}
\end{figure*}

Due to the unknown nature of the cocoon energy $E_c$, its Lorentz
factor $\Gamma_{\infty,c}$, its geometry, and the equipartition
parameters of the external shock $\epsilon_e$ and $\epsilon_B$, and
the interstellar density $n_{\rm{ISM}}$, we have explored cocoon
afterglows for a range of all those parameters. In particular, we
explore afterglows from cocoon with $E_c=10^{48}$, $10^{49}$, and
$10^{50}$ erg; Lorentz factor $\Gamma_{\infty,c}=2$, 5, and 10;
$\epsilon_e=0.01$, 0.03, and 0.1; $\epsilon_B=10^{-4}$, $10^{-3}$, and
$10^{-2}$; and $n_{\rm{ISM}}=0.001$, 0.01, and 0.1~cm${-3}$. For each
combinations of the above parameters, the X-ray, optical, and radio
afterglow are computed. The peak fluxes for the three bands are shown
in Figure~\ref{fig:new} versus the time of the peak. It is obvious
that the uncertainty in the cocoon physics and external shock
properties cause significant spread in the prediction. The brightest
transients, however, always peak a few hours after the burst in the
X-ray and optical, and within $\sim$a week in the radio.

\section{Summary and Discussion}

We have calculated the on- and off-axis emission of a typical short
GRB. Our calculations include the prompt and afterglow emission from
the relativistic jet material as well as the prompt and afterglow
emission from the cocoon material. An energetic cocoon is expected to
form as the SGRB relativistic jet propagates through the immediate
surroundings of the NS binary merger, polluted with
$\sim0.01$~M$_\odot$ of material tidally ejected from the merging
compact objects
\citep{Kiuchietal2014,Kiuchietal2015,Radiceetal2016}. We calculated
the energy of the cocoon and find it to amount to approximately $10$
per cent of the burst energy and to be a strong function of the size
of the high-density region surrounding the merger site.

As reported in Figures ~\ref{fig:xray},~\ref{fig:optical},
and~\ref{fig:radio}, the emission from the relativistic jet (both
prompt and afterglow) dominate at all times and at all wavelengths for
the on-axis observer. However, as discussed above, the most likely
observer angles are large, half of the events been observed at 35
degrees or more. In such cases (Figure~\ref{fig:coc_prompt} and the
last three panels of Figures~\ref{fig:xray},~\ref{fig:optical},
and~\ref{fig:radio}) the jet emission is undetectable with current
instrumentation, with the exception of the radio band, in which a
100~$\mu$Jy source would be detectable approximately one year after
the GW detection. A faint optical transient peaking $\sim2$~months
after the GW trigger would also be detectable for observers at
$\theta_{\rm{obs}}<50^\circ$. This is due to the fact that the
emission from an off-axis jet is dramatically reduced by relativistic
beaming (see Figure~\ref{fig:prompt}), and the jet emission is
detected only after the jet has slowed down to trans-relativistic
speed, about one year after the merger. The cocoon afterglow, instead,
is isotropic and peaks at a few hours (X-rays and optical) to a few
days (radio) after the merger. At peak time, and for a few days to a
few weeks, it is faint but clearly detectable by current
instrumentation.  However, the faint multiwavelength afterglow would
require previous localization in order to be observable with
narrow-field instruments. The faint but detectable X-ray thermal pulse
of the prompt cocoon emission gives therefore the best hope of
localizing the EM counterpart of a GW-detected binary NS merger (see
Figure~\ref{fig:coc_prompt}).

We therefore conclude that, should a GW detected NS binary merger be
promptly localized, rapid follow up would be able to detect the cocoon
afterglow emission and allow for the identification of the
electromagnetic counterpart of the GW source. Besides being brighter,
the cocoon emission is detectable just a few hours after the GW
signal, greatly reducing the likelihood of a false detection triggered
by an unrelated transient within the error radius of the GW
source. The cocoon also produces a short ($\sim0.3$~s) pulse of prompt
emission with a broadened thermal spectrum. According to our
calculations, the cocoon prompt emission is just above the detection
threshold of both the Swift BAT and the Fermi GBM.

Our calculations are based on a set of simplifying assumptions. First,
we assume that the relativistic outflow is a top-hat jet, with a sharp
edge in which the Lorentz factor drops from 100 to 1 with no boundary
layer. The jet is assumed not to spread laterally. However, sideways
expansion does not change significantly the afterglow luminosity, even
for off-axis observers (compare Figures~6 and~7 of
\cite{Rossietal2004}). We also assume that the cocoon is perfectly
isotropic, and that some degree of mixing with the ambient medium
causes the entropy of the cocoon to be lower than the one of the jet
material. While these assumption allow us to perform the calculations
using semi-analytical techniques, they might overemphasize the
differences between the on-axis and off-axis observers (see, e.g., the
results of the simulations of \cite{Gottliebetal2017}). For example, a
jet with a transition layer with lower Lorentz factor would produce a
smoother decline of the prompt emission with observer angle (compared
to the dramatic decrease seen in Figure~\ref{fig:prompt}). In
addition, a cocoon that maintains some degree of asymmetry would be
dimmer -- and therefore harder to detect -- for a binary merger with a
very large viewing angle. In Figure~\ref{fig:new} we show the
consequence of changing the values of ur fiducial parameters on the
afterglow fluxes in the X-ray, optical, and radio bands. Finally, we
neglect the effect of X-ray scattering that could add a
quasi-isotropic component to the SGRB emission
\citep{Kisakaetal2015b}.

Numerical simulations need to be performed to correctly represent the
ambient medium and self-consistently predict the jet structure and the
cocoon structure and degree of mixing with the ambient medium. We also
should note that we neglect the emission from a kilonova (also known
as a macronova), possibly associated with the merger
\citep{LiPaczynski1998,Metzgeretal2010,Kasenetal2015,Kisakaetal2015}. The
kilonova emission is isotropic and should be sufficiently bright in
the optical and IR bands ($R \sim 24$ at 200 Mpc) and peak at
$\sim1$~week after the merger. It would therefore be identified easily
from the cocoon and jet afterglows that peak on time scales of a few
hours and a few months (for a large angle observer). A kilonova
precursor peaking $\sim1$~hour after the merger has also been
discussed \citep{Metzgeretal2015}, possibly outshining the cocoon
optical afterglow component at early time.

To conclude, we would like to point out the opportunity given by the
follow-up of GW detected binary merger to pin down the jet structure
of SGRBs. When a large sample of GW detected NS binary mergers will be
available, comparing the brightness of their electromagnetic
counterparts could allow us to map the polar distribution of the jet
energy, velocity, and possibly magnetization. This would be an
extremely important constrain for jet acceleration models that is
impossible to obtain from long-duration GRBs.  However, no GW emission
from binary NS mergers has been detected so far, and using
multimessenger detections for constraining jet parameters may be a
feat that will be possible only many decades from today.

\section*{Acknowledgements}
We thank Giancarlo Ghirlanda, Edo Berger, and Brian Metzger for useful
discussions. DL acknowledges support from NASA ATP grant
NNX17AK42G. AD thanks the department of Physics of Oregon State
University for the hospitality during part of the preparation of this
paper. BJM was supported by the NSF under grant AST-1333514 and by the
Aspen Center for Physics under NSF grant PHY-1066293

\end{document}